\documentstyle[prd,aps,epsfig,twocolumn,amstex,floats]{revtex}

\def\ppbar{$p\overline{p} $}            
\def\pbarp{$\overline{p}p $}            
\def\qqbar{$q\overline{q}$}             
            \def\ttbar{$t\overline{t}$}             
\def\pt{$p_T$}                          
\def\et{$E_T$}                          
\def\met{\mbox{${\hbox{$E$\kern-0.6em\lower-.1ex\hbox{/}}}_T$}} 
\def\ipb{pb$^{-1}$}                     
\def\gevcc{GeV}                         
\def\gevc{GeV}                          
\def\gev{GeV}                           
\def\lum{$\cal{L}$}                     
\def\mw{$M_W$}                          
\def\D0{D\O}                            
\def\etal{{\sl et al.}}                   

\newcommand{\RESBO}{{\sc resbos}}
\newcommand{\LEGA}{{\sc legacy}}
\newcommand{\GEAN}{{\sc geant}}
\newcommand{\PYTH}{{\sc pythia}}

\newcommand{\HERW}{{\sc herwig}}

\newcommand{\MISE}{{\sc miser}}
\def\wb{$W$}
\def\zb{$Z$}
\def\ee{$e^+e^-$}
\def\tev{TeV}
\def\Dzero{D\O}
\def\dsdpt{$d\sigma/dp_T$}
\def\smdsdpt{$d\sigma^\prime/dp_T$}
\def\ie{{\sl i.e.}}
\def\eg{{e.g.}}
\def\gevsq{GeV$^2$}
\def\gevhalf{GeV$^{1/2}$}
\def\zbee{$Z\rightarrow ee$}
\def\wben{$W\rightarrow e\nu$}

\def\pte{$p_T^e$}
\def\vecpte{$\vec{p}_T^e$}
\def\ptee{$p_T^{ee}$}
\def\pten{$p_T^{e\nu}$}

\def\ptW{$p_T^W$}
\def\px{$p_x$}
\def\py{$p_y$}
\def\pz{$p_z$}
\def\upar{{$u_{||}$}}
\def\ttbar{{$t\overline{t}$}}
\def\zbtt{{$Z\rightarrow \tau^+\tau^-$}}

\begin{document}

\onecolumn
\title{ Measurement of the inclusive differential cross section for 
        \zb\ bosons as a function of transverse momentum in
        \pbarp\ collisions at $\sqrt{s}=1.8$ TeV}

%
\author{                                                                      
B.~Abbott,$^{45}$                                                             
M.~Abolins,$^{42}$                                                            
V.~Abramov,$^{18}$                                                            
B.S.~Acharya,$^{11}$                                                          
I.~Adam,$^{44}$                                                               
D.L.~Adams,$^{54}$                                                            
M.~Adams,$^{28}$                                                              
S.~Ahn,$^{27}$                                                                
V.~Akimov,$^{16}$                                                             
G.A.~Alves,$^{2}$
% ...
\\                                                                            
\vskip 0.30cm                                                                 
\centerline{(D\O\ Collaboration)}                                             
\vskip 0.30cm                                                                 
}                                                                             
\address{                                                                     
\centerline{$^{1}$Universidad de Buenos Aires, Buenos Aires, Argentina}       
\centerline{$^{2}$LAFEX, Centro Brasileiro de Pesquisas F{\'\i}sicas,         
                  Rio de Janeiro, Brazil}                                     
\centerline{$^{3}$Universidade do Estado do Rio de Janeiro,                   
                  Rio de Janeiro, Brazil}                                     
\centerline{$^{4}$Institute of High Energy Physics, Beijing,                  
                  People's Republic of China}                                 
\centerline{$^{5}$Universidad de los Andes, Bogot\'{a}, Colombia}             
\centerline{$^{6}$Universidad San Francisco de Quito, Quito, Ecuador}         
\centerline{$^{7}$Institut des Sciences Nucl\'eaires, IN2P3-CNRS,             
                  Universite de Grenoble 1, Grenoble, France}                 
\centerline{$^{8}$DAPNIA/Service de Physique des Particules, CEA, Saclay,     
                  France}                                                     
\centerline{$^{9}$Panjab University, Chandigarh, India}                       
\centerline{$^{10}$Delhi University, Delhi, India}                            
\centerline{$^{11}$Tata Institute of Fundamental Research, Mumbai, India}     
\centerline{$^{12}$Kyungsung University, Pusan, Korea}                        
\centerline{$^{13}$Seoul National University, Seoul, Korea}                   
\centerline{$^{14}$CINVESTAV, Mexico City, Mexico}                            
\centerline{$^{15}$Institute of Nuclear Physics, Krak\'ow, Poland}            
\centerline{$^{16}$Institute for Theoretical and Experimental Physics,        
                   Moscow, Russia}                                            
\centerline{$^{17}$Moscow State University, Moscow, Russia}                   
\centerline{$^{18}$Institute for High Energy Physics, Protvino, Russia}       
\centerline{$^{19}$Lancaster University, Lancaster, United Kingdom}           
\centerline{$^{20}$University of Arizona, Tucson, Arizona 85721}              
\centerline{$^{21}$Lawrence Berkeley National Laboratory and University of    
                   California, Berkeley, California 94720}                    
\centerline{$^{22}$University of California, Davis, California 95616}         
\centerline{$^{23}$University of California, Irvine, California 92697}        
\centerline{$^{24}$University of California, Riverside, California 92521}     
\centerline{$^{25}$Florida State University, Tallahassee, Florida 32306}      
\centerline{$^{26}$University of Hawaii, Honolulu, Hawaii 96822}              
\centerline{$^{27}$Fermi National Accelerator Laboratory, Batavia,            
                   Illinois 60510}                                            
\centerline{$^{28}$University of Illinois at Chicago, Chicago,                
                   Illinois 60607}                                            
\centerline{$^{29}$Northern Illinois University, DeKalb, Illinois 60115}      
\centerline{$^{30}$Northwestern University, Evanston, Illinois 60208}         
\centerline{$^{31}$Indiana University, Bloomington, Indiana 47405}            
\centerline{$^{32}$University of Notre Dame, Notre Dame, Indiana 46556}       
\centerline{$^{33}$Purdue University, West Lafayette, Indiana 47907}          
\centerline{$^{34}$Iowa State University, Ames, Iowa 50011}                   
\centerline{$^{35}$University of Kansas, Lawrence, Kansas 66045}              
\centerline{$^{36}$Kansas State University, Manhattan, Kansas 66506}          
\centerline{$^{37}$Louisiana Tech University, Ruston, Louisiana 71272}        
\centerline{$^{38}$University of Maryland, College Park, Maryland 20742}      
\centerline{$^{39}$Boston University, Boston, Massachusetts 02215}            
\centerline{$^{40}$Northeastern University, Boston, Massachusetts 02115}      
\centerline{$^{41}$University of Michigan, Ann Arbor, Michigan 48109}         
\centerline{$^{42}$Michigan State University, East Lansing, Michigan 48824}   
\centerline{$^{43}$University of Nebraska, Lincoln, Nebraska 68588}           
\centerline{$^{44}$Columbia University, New York, New York 10027}             
\centerline{$^{45}$New York University, New York, New York 10003}             
\centerline{$^{46}$University of Rochester, Rochester, New York 14627}        
\centerline{$^{47}$State University of New York, Stony Brook,                 
                   New York 11794}                                            
\centerline{$^{48}$Brookhaven National Laboratory, Upton, New York 11973}     
\centerline{$^{49}$Langston University, Langston, Oklahoma 73050}             
\centerline{$^{50}$University of Oklahoma, Norman, Oklahoma 73019}            
\centerline{$^{51}$Brown University, Providence, Rhode Island 02912}          
\centerline{$^{52}$University of Texas, Arlington, Texas 76019}               
\centerline{$^{53}$Texas A\&M University, College Station, Texas 77843}       
\centerline{$^{54}$Rice University, Houston, Texas 77005}                     
}                                                                             

\maketitle

\begin{abstract}
We present a measurement of the differential cross section
as a function of transverse momentum
of the \zb\ boson in \ppbar\ collisions at
$\sqrt{s}=1.8$ \tev\ using data collected by the \Dzero\ experiment
at the Fermilab Tevatron Collider during 1994--1996. We find good
agreement between our data and the NNLO resummation prediction and
extract values of the non-perturbative parameters for the resummed
prediction from a fit to the differential cross section.
\end{abstract}
%
%
\twocolumn


\clearpage

\section{Introduction}
\label{sec-introduction}
The study of the production properties of the
\zb\ boson began in 1983 with its discovery by the UA1 and UA2
collaborations at the CERN \ppbar\
collider~\cite{UA1_Z_discovery,UA2_Z_discovery}. Together with the
discovery of the \wb\ boson~\cite{UA1_W_discovery,UA2_W_discovery} 
earlier that year, the observation of the
\zb\ boson provided a direct confirmation of the unified model of
the weak and electromagnetic interactions, which, together with
QCD, is now called the standard model. Since its discovery, many of
the intrinsic properties of the \zb\ boson have been examined in
great detail via \ee\ collisions at the LEP \ee\ collider at CERN
~\cite{LEPZreview}. The mass of the \zb\ boson measured at LEP and
the SLC \ee\ collider at SLAC, known to better than one part in
$10^4$~\cite{PDB}, is one of the most
precisely measured parameters in particle physics.

LEP experiments have focused on the intrinsic properties of the
\zb\ boson, examining the electroweak character of its production
and decay in \ee\ collisions. At the Tevatron, where the \zb\ boson
is produced in \ppbar\ collisions, its production properties are
presumably characterized by QCD. Since the electroweak properties
of the \zb\ boson are not correlated with the strong properties of its
production, the \zb\ boson can therefore serve as a
clean probe of the strong interaction. Also, the large
mass of the \zb\ boson assures a large energy scale ($Q^2\approx
M_Z^2)$ for probing perturbative QCD with good reliability. The
measurement of the cross section as a function of transverse
momentum (\dsdpt) of the
\zb\ boson provides a sensitive test of QCD at high-$Q^2$. In this
article, we describe a measurement of
\dsdpt\ of the \zb\ boson using the \ee\  decays of the 
\zb ~\cite{Casey_thesis}.

In the parton model, at lowest order, \zb\ bosons are produced in
head-on collisions of \qqbar\ constituents of the proton and
antiproton, and cannot have any transverse momentum. Consequently,
the fact that observed \zb\ bosons have finite \pt\ is attributed
to gluon radiation from the colliding partons prior to their
annihilation into the \zb\ boson. Gluon radiation within the color
field of the proton or antiproton increases in proportion to the
time available for such annihilation, which is proportional to the 
inverse of the energy scale 
for the process ($1/Q$)~\cite{CSS}. The radiated gluons carry away
transverse momentum from the annihilating quarks and momentum
conservation requires that this be observed in the \pt\ of the
\zb\ boson. Thus, one expects that the observed transverse momentum
distribution of any dielectron system (produced at a scale $Q\approx
M_{ee}$) will broaden as a function of $Q$. This is, indeed, the
effect observed. At $M_{ee}\approx10$ \gevcc, the typical
\pt\ for Drell-Yan pairs\cite{drellyan} is about 1 \gevc\cite{DYexpt}.
For \wb\ boson production ($Q\approx80$ \gevcc), the average \pt\
is about 5
\gevc\cite{Wptexpt}. For \zb\ boson production ($Q\approx91$ \gevcc), the
average \pt\ is about 6 \gevc \cite{CDF_ptz}.

In general, the differential cross section for producing the state
$V$ is given by:
\begin{equation}\label{eq:diffxsec1}
  {d^2\sigma_{ij\rightarrow V}\over dp_T^2dy}=
   \sum_{i,j}\int dx_i dx_j f(x_i)f(x_j)
  {d^2\hat\sigma_{ij\rightarrow V}\over dp_T^2dy}
\end{equation}
where $p_T$ and $y$ are the transverse momentum and the rapidity of
the state $V$; $x_i$ and $x_j$ are the momentum fractions of the
colliding partons; $f(x_i)$ and $f(x_j)$ are the parton
distribution functions (pdf's) for the incoming partons; and
$\hat\sigma_{ij\rightarrow V}$ is the partonic cross section for
production of the state V, in our case, the \zb\ boson. The
subscripts $i$ and $j$ denote the contributing parton flavors (\ie,
up, down, etc.) and the sum is over all such flavors.

In standard perturbative QCD (pQCD), one calculates the partonic
cross section by expanding in powers of the strong coupling
constant, $\alpha_s$. This procedure works well when $p_T^2\sim
Q^2$. However, as $p_T\rightarrow 0$, correction terms that are
proportional to $\alpha_s\ln(Q^2/{p_T^2})$ become significant for
all values of $\alpha_s$, and the cross section diverges at small
\pt. Physically, the failure of the calculation is due to the
presence of collinear and low-\pt\ gluons that are not properly
accounted for in the standard perturbative expansion.  This
difficulty is surmounted by reordering the perturbative series
through a technique called {\it resummation}
\cite{CSS,DaviesStirling,AEGM,DWS,ArnoldReno,AK,LadinskyYuan,BalazsYuan}.

In its final form, the differential cross section is calculated as a
Fourier transform in impact parameter, $b$, space:
\begin{equation}
        \label{eq:diffxsect2}
        {d^2\sigma_{ij\rightarrow V}\over d{p_T^2}dy}\approx
        \int_0^\infty d^2b\ e^{i\vec{p}_T\cdot\vec{b}}W(b,Q)+Y(b,Q)
\end{equation}
where $W(b,Q)$ contains the results of resumming the perturbative
series, and $Y(b,Q)$ adds back to the calculation the pieces that
are perturbative in $\alpha_s$, but are not singular at
$p_T=0$\cite{CSS}.

Although the resummation technique extends the applicability of
pQCD to lower values of $p_T$, a more fundamental barrier is
encountered when \pt\ approaches $ \Lambda_{\text QCD}$, and pQCD is
expected to fail in general. In this region, we expect
non-perturbative aspects of the strong force to dominate the
production of the vector boson. This implies that $W(b,Q^2)$ in
Eq.~\ref{eq:diffxsect2} is undefined above some value of
$b=b_{max}$. To extend the calculation to $p_T = 0$, the
following substitution is made:
\begin{equation}
        \label{eq:wbstar}
        W(b,Q)\rightarrow W(b_*,Q)e^{-S_{NP}(b,Q)}
\end{equation}
where $b_*\equiv b/\sqrt{1+(b/b_{\rm max})^2}$. This effectively cuts off
the contribution of $W(b,Q)$ near $b_{\rm max}$, leaving the
differential cross section dominated by $S_{NP}$, where
$S_{NP}(b,Q)$ is called the {\it non-perturbative Sudakov form
factor}. $S_{NP}$ has the generic renormalization group invariant
form\cite{CSS}
\begin{eqnarray}
        \label{eq:snpgen}
        S_{NP}(b,Q)= \nonumber \hspace{-1cm}\\
        & h_1(b,x_i)+h_1(b,x_j)+h_2(b)\cdot\ln({Q \over 2Q_o})
\end{eqnarray}
where $x_i$ and $x_j$ are the momentum fractions of the
annihilating quarks; $Q_o$ is an arbitrary momentum scale; and
$h_1(b,x)$, $h_2(b)$ are phenomenological
functions to be determined from
experiment\cite{DWS,AK,LadinskyYuan}. The fact that $h_2(b)$ lacks
any dependence on the momentum fractions of the incoming partons has
led to speculation that it may contain some deeper relevance to the
gluonic structure of the proton\cite{KorchSterm}.

The current understanding of the \pt\ distribution for \zb\ bosons
uses fixed-order perturbative calculations [leading-order (LO) or
next-to-leading-order (NLO)] to describe the high-\pt\ region, and
resummation calculations of the perturbative solution to describe
the low-\pt\ region. The resummation calculation fails at large-\pt\ 
(of the order 50 \gevc) due to large terms missing from the calculation 
resulting from the $p_T\rightarrow 0$ approximation. An {\it ad hoc} 
``matching'' criteria is  invoked to decide when to switch from the 
resummed calculation to  the fixed-order calculation, which is 
considered to be robust at  large-\pt. Additionally, a parameterization of
Eq.~\ref{eq:snpgen} is invoked to account for
non-perturbative effects at the lowest \pt\ values which are not
calculable in perturbative QCD.

In our measurement of the \pt\ distribution, we restrict the
invariant mass of the dielectron system to be approximately the
mass of the \zb\ boson, where the \zb\ resonance greatly dominates
dielectron production. The remaining contribution is due almost
entirely to production of \ee\ pairs via the photon propagator
(Drell-Yan process), which is coherent and interferes quantum
mechanically with \zb\ boson production. Other processes also contribute
to inclusive dielectron production in
\ppbar\ collisions, \eg,
\ttbar\ and diboson production, however, these are incoherent with
\zb\ boson production and their overall rate is negligibly small.

Besides being of intrinsic interest in the study of QCD, precise
understanding of \zb\ boson production in
\ppbar\ collisions has important practical benefits for other
measurements with electrons in the final state. The phenomenology
used to describe \zb\ boson production is applicable to \wb, \zb,
and essentially all Drell-Yan type processes. In the low-\pt\
region, where the cross section is highest, uncertainties in the
phenomenology of vector boson production have contributed
to the uncertainty in the measurement of the mass of
the \wb\ boson (\mw)~\cite{D0Wmass,CDFWmass}. Additionally,
diboson, top quark, and Higgs boson production all have single and
dielectron backgrounds from \wb\ and \zb\ boson production that will
be more constrained through a precise measurement of \zb\ boson
production properties.

Despite larger statistical uncertainties relative to \wb\ boson
production (there are $\sim$10 times more \wben\ than \zbee\ events
produced at $\sqrt{s}=1.8$ \tev), the \zb\ boson provides a better
laboratory for evaluating the phenomenology of vector boson
production. The measurement of the transverse momentum of the
\ee\ pair (\ptee) does not suffer from the same level of
experimental imprecision as the measurement of \pten\ because the
latter relies on the determination of the total missing transverse
momentum in the detector (\met), which has inherently higher
systematic uncertainties. The typical resolution in \ptee\ is about
1.5 \gevc\ compared with 4--5
\gevc\ for \pten, and the \ptee\ resolution is approximately flat
as \ptee\ increases, whereas it continues to degrade for \pten.

Previous measurements of the differential cross section for
\zbee\ production in \ppbar\ collisions have been limited primarily
by statistics. The UA2 \cite{UA2_ptz} collaboration analyzed 162
events, concluding that there was basic agreement with QCD, but
that more statistics were needed. In the 1988--89 run at the
Tevatron, CDF~\cite{CDF_ptz} analyzed 235 dielectron events and
103 dimuon events, making similar conclusions. 
Our study is based on a total of about 6400 events. We determine the \pt\
distribution for the \zb\ boson and use our results to constrain the 
non-perturbative Sudakov form factor. We then remove the effects of 
detector smearing  and obtain a normalized differential cross section
\dsdpt. 

We present a brief description of the \D0\ detector in the next
section. We then present the selection procedure for our data
sample. The selection efficiency (Section ~\ref{sec-efficiency}),
kinematic and fiducial acceptances (Section ~\ref{sec-acceptance}),
contributing backgrounds (Section ~\ref{sec-backgrounds}), fit for
non-perturbative parameters (Section~\ref{sec-fitgvalues}), and
the smearing correction
(Section~\ref{sec-smearcorr}) are all discussed in turn. These
individual components are combined (Section ~\ref{sec-results}) to
obtain the final differential cross section, which is compared to
predictions from QCD.

\section {Experimental Setup}
\label{sec-exper}
\begin{figure}[htpb!]
\vspace{0.5in}
\centerline{\psfig{figure=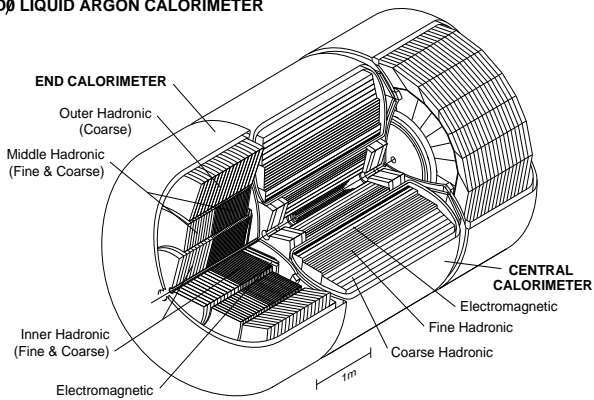,width=3.3in,height=2.5in}}
\vspace{-0.02in}
\caption{ A cutaway view of the \Dzero\ calorimeter and tracking system.}
\label{fig:d0cal}
\end{figure}

The \Dzero\ detector consists of three major subsystems: a central
detector, a calorimeter (Fig.~\ref{fig:d0cal}), and a muon
spectrometer. It is discussed in detail elsewhere~\cite{d0nim}. We
describe below only the features that are most relevant for this
measurement.

\subsection {Conventions}

We use a right-handed Cartesian coordinate system with the $z$-axis
defined by the direction of the proton beam, the $x$-axis pointing
radially out of the Tevatron ring, and the $y$-axis pointing up. A
vector $\vec p$ is then defined in terms of its projections on
these three axes, \px, \py, \pz. Since protons and antiprotons in
the Tevatron are unpolarized, all physical processes are invariant
with respect to rotations around the beam direction. It is
therefore convenient to use a cylindrical coordinate system, in
which the same vector is given by the magnitude of its component
transverse to the beam direction, \pt, its azimuth $\phi$, and \pz.
In \ppbar\ collisions the center of mass frame of the parton-parton
collisions is approximately at rest in the plane transverse to the
beam direction, but has an unknown boost along the beam direction
due to the dispersion of the parton momentum fraction within the
interacting proton and antiproton. Consequently, the total
transverse momentum vector in any event (\met) must be
close to zero, and can be used to reject background from events
that have neutrinos in the final state.
We also use spherical coordinates by replacing \pz\ with the
colatitude $\theta$ or the
pseudorapidity $\eta=-\ln\tan\left(\theta/2\right)$. The origin of
the coordinate system is, in general, defined as the reconstructed
position of the \ppbar\ interaction for describing the interaction,
and the geometrical center of the detector when describing the
detector. For convenience, we use natural units ($\hbar=c=1$)
throughout this paper.
Additionally, we use ``\pt '' to refer to the transverse momentum of
the \zb\ boson or objects which mimic the \zb, e.g., background
events in which the momenta for the objects considered to form the 
fake-\zb\ are added together to generate a \pt\ value. Deviations 
will be noted with an appropriate superscript. 

\subsection{Central Detector}
The central detector is designed to measure the trajectories of
charged particles. It consists of a vertex drift chamber, a
transition radiation detector, a central drift chamber (CDC), and
two forward drift chambers (FDC). There is no central magnetic
field, and \Dzero\ therefore cannot distinguish particles by their
electric charge, with the exception of muons which penetrate the outer
toroidal magnets. Consequently, in the rest of this paper, the term
electron will refer to either an electron or a positron.
The CDC covers the detector pseudorapidity region
$|\eta_{\rm det}|<1.0$.
It is a drift chamber with delay lines that give the hit
coordinates along the beam direction ($z$) and transverse to the
beam ($r$, $\phi$). 
The FDC covers the region $1.4<|\eta_{\rm det}|<3.0$.

\subsection{Calorimeter}
\label{sec-exper-cal}

The sampling calorimetry is contained in three cryostats, each
primarily using uranium absorber plates and liquid argon as the
active medium. There is a central calorimeter (CC) and two end
calorimeters (EC). Each is segmented into electromagnetic (EM)
sections, a fine hadronic (FH) section, and coarse hadronic (CH)
sections, with increasingly coarser sampling. The entire
calorimeter is divided into about 5000 pseudo-projective towers,
each covering 0.1$\times$0.1 in $\eta\times\phi$. The EM section is
segmented into four layers which are 2, 2, 7, and 10 radiation lengths
in depth respectively. The third layer, in which electromagnetic
showers reach their maximum energy deposition, is further
segmented into cells covering 0.05$\times$0.05 in $\eta\times\phi$.
The hadronic sections are segmented into four (CC) or five (EC)
layers. The entire calorimeter is 7--9 nuclear interaction lengths
thick. There are no projective cracks in the calorimeter, and it
provides hermetic and nearly uniform coverage for particles with $|
\eta_{\rm det}|<4$.

\subsection{Trigger}
Readout of the detector is controlled by a multi-level trigger
system. The lowest level hardware trigger consists of two arrays of
scintillator hodoscopes, which register hits with a 220 ps time
resolution and are mounted in front of the EC cryostats. Particles
from the breakup of the proton and the antiproton produce hits
in hodoscopes at opposite ends of the CC, each of which are tightly
clustered in time. At the lowest trigger level, the detector has a
98.6\%\ acceptance for \wb/\zb\ boson production.
For events that contain only a single \ppbar\ interaction, the
location of the interaction vertex can be
determined from the time difference between the hits at the two
ends of the detector to an accuracy of 3 cm. This interaction vertex
is used in the last level of the trigger.

The next trigger level consists of an AND-OR decision network
programmed to trigger on a
\ppbar\ crossing when several preselected conditions are met. This
decision is made within the 3.5 $\mu$s time interval between beam
bunch crossings. The signals from 2$\times$2 arrays of calorimeter
towers (``trigger towers"), covering 0.2$\times$0.2 in
$\eta\times\phi$, are added together electronically for the EM
sections (``EM trigger towers'') as well as for all sections,
and shaped with a fast rise
time for use at this trigger level. An additional trigger processor
can be invoked to execute simple algorithms on the limited
information available at the time of the AND-OR network. These
algorithms use the energy deposits in each of the calorimeter trigger
towers.

The final software-based level of the trigger consists of an array
of 48 VAXstation 4000 computers. At this level, complete event
information is available and more sophisticated algorithms are
used to refine the trigger decisions. Events are accepted based
on certain preprogrammed conditions and  are recorded for eventual
offline reconstruction.

\section{Data Selection}
\label{sec-data_selection_and_reconstruction}
\subsection{Trigger Filter Requirements}

We require the transverse energy, \et\ ($E\sin\theta$), of one 
or more trigger towers
to be greater than 10 \gev. The trigger processor computes an EM
transverse energy by combining the \et\ of the EM trigger tower
(that exceeded some threshold) with the largest signal in the
adjacent EM trigger towers, but doing this only if the original EM
signal has at least 85\% of the energy of the entire trigger tower
(including hadronic layers).

For the accepted trigger tower, a software algorithm finds the
most energetic of four sub-towers, and sums the energy in a
3$\times$3 array of calorimeter cells around it. It examines the
longitudinal shower profile by checking the fraction of the total 
energy found in different EM layers. The transverse shower shape is
characterized by the pattern of energy deposition in the third EM
layer. The difference between the energies in concentric regions
centered on the most energetic tower covering 0.25$\times$0.25 and
0.15$\times$0.15 in $\eta\times\phi$ must be consistent with
expectations for an electron shower. The trigger also imposes an
isolation condition requiring
\begin{eqnarray}
 \frac{\sum_iE_i\sin\theta_i - p_T^e}{p_T^e} < 0.15
\end{eqnarray}
where the sum runs over all cells within a cone of radius
${\cal R}=\sqrt{\Delta\phi^2+\Delta\eta^2}=0.4$ around the electron
direction and \pte\ is the transverse momentum of the electron,
based on its energy and the $z$-position of the interaction vertex
as measured by the hodoscopes.

\begin{figure}[htpb!]
\hspace{3.3cm}
\centerline{\psfig{figure=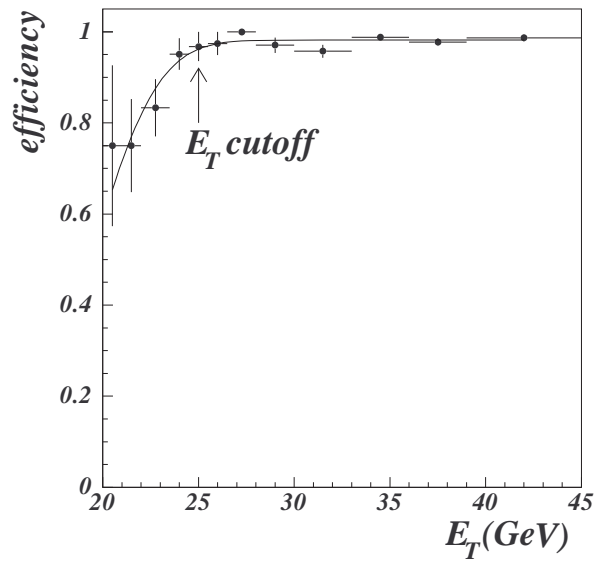,width=3.2in}}
\caption{Electron detection efficiency as a function of electron \et\ at
the trigger level. The efficiency is essentially flat above our
final \et\ cutoff of $25$ \gevc.}
\label{fig:ptetrig}
\end{figure}

The trigger requires two electrons which satisfy the isolation
requirement, each with \et$>$20 \gev. Figure~\ref{fig:ptetrig}
shows the measured detection efficiency of the electron filter as a
function of \pte\ for a threshold of 20 \gevc. We determine
this efficiency using \zb\ boson data taken with a lower threshold value
(16 \gevc). (The efficiency corresponds to the fraction of
electrons found at the higher threshold.) The curve is a
parameterization used in the simulation described in Section
\ref{sec-cms}.

\subsection{Fiducial and Kinematic Requirements}

Events passing the filter requirement are analyzed offline where
they are reconstructed with finer precision. The two highest-\et\
electron candidates in the event, both having \et $>$ 25 \gev, are
used to reconstruct the \zb\ boson candidate. One electron is required 
to be in the central region, $|\eta_{\rm det}|<1.1$ (CC), and the second
electron may be either in the central or in the forward region,
$1.5<|\eta_{\rm det}|<2.5$ (EC). This yields two topologies for the
selected events: CCCC, where both electrons are detected in the
central region, and CCEC, where one electron is detected in the
central region and the other in the forward region. In order to
avoid areas of reduced response between neighboring $\phi$ modules
of the central calorimeter, the $\phi$ of any electron is required
to be at least $0.05 \times 2 \pi / 32$ radians away from the 
position of a module
boundary. Finally, the events are required to have an invariant
mass near the known value of the \zb\ boson mass, $75 < M_{ee}< 105 $
\gevcc.

\subsection{Electron Quality Criteria}

To be acceptable candidates for \zb\ production, both
electrons are required to be isolated and to satisfy offline
cluster-shape requirements. Additionally, at least one of the
electrons is required to have a spatially matching track associated 
with the reconstructed calorimeter cluster.

The isolation fraction is defined as
\begin{equation}
f_{\rm iso}={E_{\rm cone}-E_{\rm core}\over E_{\rm core}},
\end{equation}
where $E_{\rm cone}$ is the energy in a cone of radius 
${\cal R}=0.4$ around
the direction of the electron, summed over the entire depth of the
calorimeters, and $E_{\rm core}$ is the energy in a cone of 
${\cal R}=0.2$,
summed over only the EM calorimeter. Both electrons in the data
sample are required to have $f_{\rm iso}<0.15$.

We test how well the shape of any cluster agrees with that expected
for an electromagnetic shower by computing the quality variable
$\chi^2_{HM}$ for all cell energies using a 41-dimensional
covariance matrix called the H-matrix\cite{hmatrix}. The covariance
matrix is determined from \GEAN-based
simulations~\cite{geant,D0top}, which were tuned to agree with test
beam measurements. Both electrons in the sample are
required to have a tight selection of $\chi^2_{HM} < 100$.

The quality of the spatial match between a reconstructed track and an
electromagnetic cluster is defined by the variable
\begin{equation}
\sigma^2 = \left({\Delta s\over \delta s}\right)^2 +
\left({\Delta z\over \delta z}\right)^2,
\end{equation}
where $\Delta s$ is the distance between the centroid of the cluster
in the third EM layer and the extrapolated trajectory of the track
along the azimuthal direction, and $\Delta z$ is the analogous
distance in the $z$-direction. For EC electrons, $z$ is replaced by
$r$, the radial distance from the center of the detector. The
parameters $\delta s = 0.25$ cm, $\delta z = 2.1$ cm, and $\delta r =
1.0$ cm, are the resolutions in $\Delta s$, $\Delta z$, and $\Delta
r$, respectively. At least one of the candidate electrons is required
to have $\sigma < 5$ for candidates with $|\eta_{\rm det}|<1.1$ and
$\sigma<10$ for candidates with $1.5<|\eta_{\rm det}|<2.5$.

\begin{figure}[htpb!]
\centerline{\psfig{figure=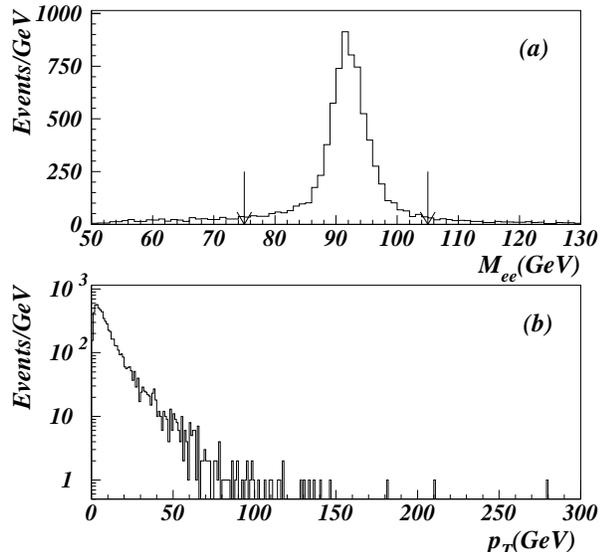,width=3.2in}}
\caption{(a) Mass distribution for all accepted electron pairs,
and (b) the \pt\ distribution for those pairs with
$75 < M_{ee}< 105 $ \gevcc.}
\label{fig:masspt}
\end{figure}

The total integrated luminosity of the data sample is 111 \ipb.
After applying the selection criteria, 6407 events remain, with 3594
events containing both electrons in the central region and 2813 events
containing one electron in the central region and one in the forward
region. Figure~\ref{fig:masspt} shows the mass and \pt\ distributions
(for $75 < M_{ee}< 105 $ \gevcc) in the final data sample. There are
157 events with \pt$>$50
\gevc, and the event with the largest \pt\ has  $p_T=280$ \gevc.

\subsection{Resolutions and Modeling of the Detector}
\label{sec-cms}
Both the acceptance and the resolution-smeared theory are
calculated using a simulation technique originally developed for
measuring the mass of the \wb\ boson~\cite{D0Wmass} and inclusive 
cross sections
of the \wb\ and \zb\ bosons~\cite{wzcross}, with minor
differences arising from small differences in the selection
criteria. We briefly summarize the simulation here.

The mass of the \zb\ boson is generated according to an energy-dependent 
Breit-Wigner lineshape. The \pt\ and rapidity ($y$) are chosen randomly
from grids created with the computer program
\LEGA~\cite{BalazsYuan} which calculates the \zb\ boson cross section 
for a given \pt, $y$, and $Q$. For calculating the grids, we use a 
fixed value for the mass of the \zb\ boson of 91.184 \gev.
 We match the low-\pt\ and high-\pt\ regions
following the algorithm used in the program
\RESBO~\cite{BalazsYuan} to produce a grid of \pt\ and $y$
values, weighted by the production cross section, calculated to
NNLO. The primary vertex distribution for the event is modeled as a
Gaussian with a width of 27 cm and a mean of
$-0.6$ cm, corresponding to the width and offset measured in the
data. The positions and energies of the electrons are smeared
according to the measured resolutions and corrected for offsets in
energy scale caused by the underlying event and recoil particles
emitted into the calorimeter towers. Underlying events are modeled
using data from random inelastic \ppbar\ collisions with the same
luminosity profile as the \zb\ sample.

\begin{figure}[htpb!]
\centerline{\psfig{figure=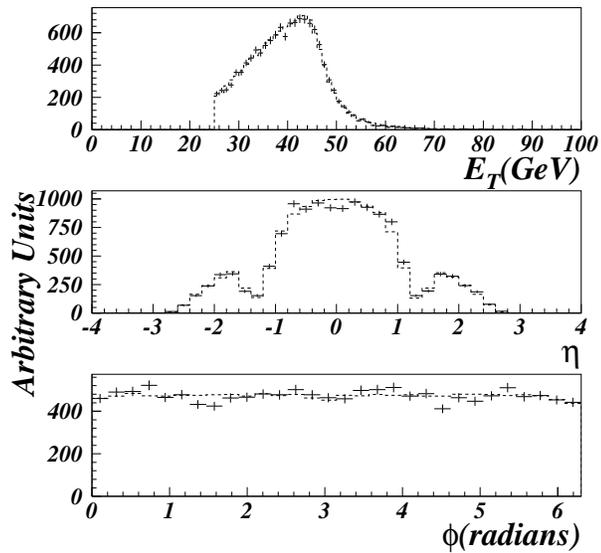,width=3.2in}}
\caption{Comparison of electron \et, $\eta$, and $\phi$, from
\zb\ boson data (crosses) to results of the detector
 simulation (dashed).}
\label{fig:mcdatacompare}
\end{figure}

The electron energy and angular resolutions are tuned to reproduce
the observed width of the \zbee\ mass distribution at the
\zb\ resonance. The fractional energy resolution can be parameterized as
a function of electron energy as $\Delta E / E = {\cal C}\oplus
{\cal S} / \sqrt{E_T}$. The sampling term, $\cal{S}$, was
obtained from measurements made in a calibration beam, and is 0.135
\gevhalf\ for the CC and 0.157 \gevhalf\ for the EC
~\cite{Zhu_thesis,Heuring_thesis}. The constant term, $\cal{C}$ was
determined specifically for our selection criteria. In the CC, the
value is ${\cal  C}=0.014 \pm 0.002$ and in the EC the value is
${\cal  C}=0.0^{+0.01}_{-0.00}$. The uncertainty is dominated by the
statistics of the \zbee\ sample. The uncertainty in the polar angle
of the electrons is parameterized in terms of the uncertainty in the
center-of-gravity of the track used to determine the polar angle.
Figure ~\ref{fig:mcdatacompare} compares electrons from \zb\ boson data with
simulated results for distributions in electron \et, pseudorapidity, 
and $\phi$.

In addition to the smearing of the electron energies and positions,
certain specific features of the experiment are also modeled in the
simulation in order to more closely represent the data. A
parameterization of the rise in efficiency of the trigger
as a function of electron \et\ is included, as well as a
parameterization of the tracking efficiency as a function of
electron pseudorapidity. Both efficiencies have a negligible effect
on the shape of the \pt\ distribution. Details of the detector
simulation can be found in
Refs.~\cite{Adam_thesis,Flattum_thesis}.

\section{Efficiency}
\label{sec-efficiency}
We determine the efficiency of the event selection criteria as a
function of the \pt\ of the \zb\ boson, normalizing the result to the
integrated total cross section for \zb\ boson production as
measured at \D0\ ($\sigma_Z\cdot B(Z\rightarrow ee)=221$
pb)\cite{wzcross}.

Of all the selection criteria, the electron isolation
requirement has the largest impact on the observed
\pt\ of the \zb\ boson. Nearby jet activity spoils the isolation of
an electron, causing it to fail the selection criteria. The effect
depends upon the detailed kinematics of the event, in particular,
the location of hadronic activity (e.g., associated jet production)
and the \pt\ of the vector boson.

\begin{figure}[htpb!]
\centerline{\psfig{figure=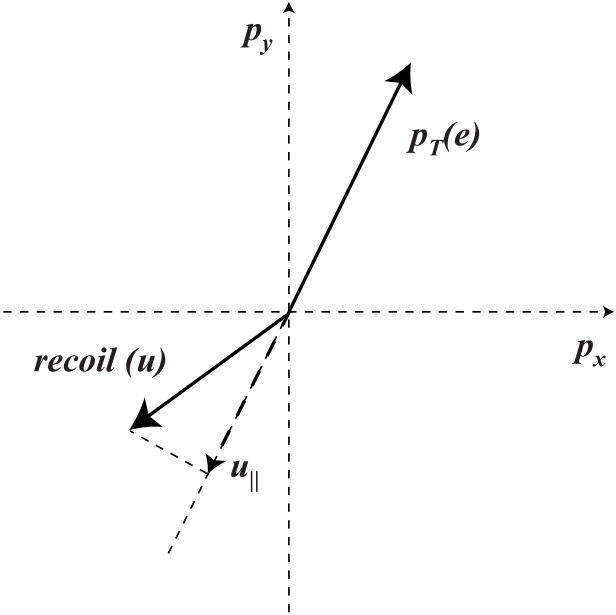,width=3.2in}}
\caption{Illustration of the relationship between the transverse
momentum of the electron, the vector \et\ of the hadron recoil ($u$) in
the calorimeter, and \upar, the projection of the recoil onto the
transverse direction of the electron. In the particular example
illustrated here, \upar\ is negative.}
\label{fig:upardiag}
\end{figure}

Two methods have been used to determine the \pt-dependence of the
electron identification efficiency. In the first method, the effect 
of jet activity near an electron shower is parameterized in terms of the
component of the hadronic recoil energy ($u$) projected onto the vector
\pte. This is denoted as \upar\ \cite{CDF_Wmass_Run0}.
The relationship between \vecpte\ and
\upar\ is illustrated in Fig.~\ref{fig:upardiag}. We used a
combination of simulated electrons and \wb\ boson data to obtain the
efficiency for identifying electrons as a function of \upar.
Electron showers were generated using the
\GEAN\ detector-simulation program, and the parameters for
the simulated electrons (e.g., \et, isolation, $\chi^2_{HM}$)
agreed well with those observed in \wb\ boson data\cite{D0Wmass}. The
agreement suggests that the effect of hadronic activity on the
electron is well-modeled in the simulation. Although our
parameterization is obtained using electrons from \wb\ events, we
apply it to electrons from \zb\ boson events (which have very similar
energy distributions due to hadronic recoil),
because the parameterization reflects the effect of hadronic
activity on high-\pt\ electrons, regardless of the origin of that
activity.

\begin{figure}[htpb!]
\centerline{\psfig{figure=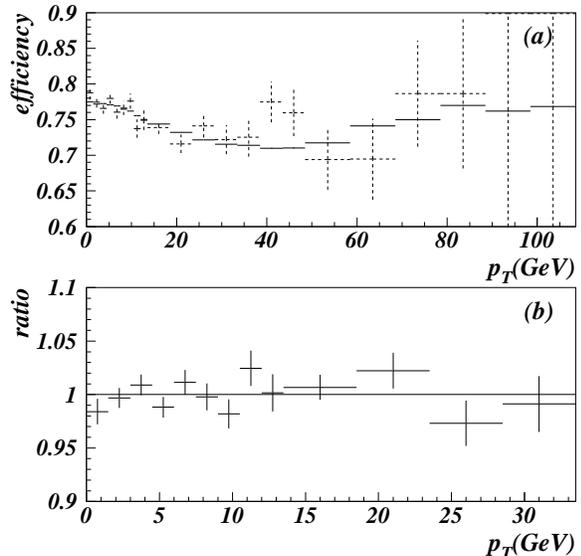,width=3.2in}}
\caption{(a) Comparison of the \zb\ boson selection efficiency
as a function of \pt\ as determined using \HERW\ (dashed crosses),
and as determined using a parameterization of the single-electron
efficiency as a function of \upar\ (solid crosses). (b) The ratio of the
two methods in the range 0--30 \gevc, where they are expected to agree.}
\label{fig:upareff}
\end{figure}

The electron identification efficiency as a function of
\upar\ is parameterized as:
\begin{equation}
\epsilon (u_{\parallel})=\left\{
        \begin{array}{cc}
        \alpha      &  {\rm if}\ (u_\parallel< u_o)  \\
        \alpha(1-s(u_{\parallel}-u_o)) &  {\rm if}\ (u_\parallel> u_o) \\
        \end{array}\right.
\end{equation}
where $u_o$ is the value of \upar\ at which the efficiency begins
to decrease with \upar, and $s$ is the rate of decrease. The values
obtained from the best fit are are $u_o=3.85\pm0.55$ \gev\ and
$s=0.013\pm 0.001$ \gev$^{-1}$. The parameter $\alpha$ reflects the
overall efficiency, which, as we have indicated, is obtained from a
normalization to the overall selection efficiency. The final event
efficiency as a function of \pt\ of the \zb\ boson, shown in
Fig.~\ref{fig:upareff}, is obtained from the detector simulation,
by comparing the \pt\ distribution with and without the
\upar\ correction. The final event efficiency is insensitive to the use of
different parameterizations of the \upar\ efficiency in the EC versus the 
CC. A more detailed description of the method
used to obtain the \upar\ parameterization can be found in
Ref.~\cite{D0Wmass}.

In the end, the \upar\ parameterization of the event
identification efficiency alone is unsatisfactory for application to
this measurement. In particular, that analysis required \ptW$<$30
\gevc, thereby restricting applicability to that region.
To obtain a reasonable parameterization of the electron
identification efficiency for all values of \pt, we extract the
\zb\ boson identification efficiency from events generated with
\HERW \cite{HERWIG}, smeared with the \Dzero\ detector resolutions,
and overlaid onto randomly selected \ppbar\ collisions
(``zero-bias" events). The efficiency as a function of
\pt\ is defined by the ratio of the \pt\ distribution for events
with resolution smearing and kinematic, fiducial and electron
quality requirements imposed, to that with only kinematic and
fiducial requirements. Figure~\ref{fig:upareff}a compares the
efficiency as a function of \pt\ using the \upar\ parmeterization
with that using the detector-smeared \HERW\ events. The
distributions have been normalized to each other in the region
\pt$<$30 \gevc. Figure~\ref{fig:upareff}(b) shows the ratio of the
two normalized results for \pt$<$30 \gevc. The agreement of the
\HERW\ analysis with the \upar\ analysis is taken as confirmation
of the validity of the \HERW\ result for all \pt. (The model for
the \upar\ analysis has been shown to be reliable for \pt$<$30
\gevc.)

\begin{figure}[htpb!]
\centerline{\psfig{figure=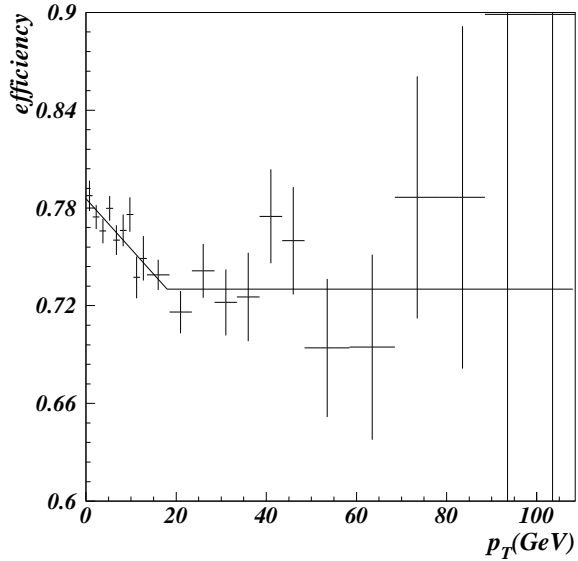,width=3.2in}}
\caption{Final event identification efficiency as a function of
\pt, based on \HERW\ events; the line is the
parameterization used in calculating the final cross section.}
\label{fig:pteff}
\end{figure}

In normalizing our efficiency to the previously determined
inclusive \zb\ boson event selection efficiency, we use the
combined CCCC and CCEC efficiency of 0.76 \cite{wzcross}. We fit
the \HERW\ result to a linear function in the region \pt$<$18
\gevc, and a constant in the region \pt$>$18 \gevc, to obtain
the \pt-dependent event selection efficiency for all \pt\ values.
The parameterization is shown in Fig. \ref{fig:pteff}. The
\pt-dependence of the efficiency, in absolute terms, is given by
$\epsilon = 0.78-0.004p_T$, for \pt$<$ 18 \gevc, and 0.73 for
\pt$>$18 \gevc.

We assume that the efficiency above 100 \gevc\ is the same as in
the region of 18--100 \gevc. This is the simplest assumption we can
make given the statistics of the simulation. The efficiency at
high-\pt\ cannot be greater than at \pt\ = 0, which would
correspond to about a 1.5 standard deviation change in the cross
section in that region, and this difference would be reflected in
the uncertainty on the extracted differential cross section. We do
not expect the efficiency to decrease in the region beyond 100
\gevc, because the jets in such events will tend to be in the
hemisphere opposite to the electrons. Events with high jet
multiplicity may have instances in which the large-\et\ jets
balance most of the transverse momentum of the event, but smaller-\et\
jets can overlap with one of the electrons. However, because the
electrons are very energetic, low energy jets are not likely to
affect the efficiency of the isolation criteria. We assign
estimated uncertainties on the efficiency of $\pm 3$\% in the
bin below 18 \gevc, and $\pm 5$\% in the region above 18 \gevc.

%
\section{Acceptance}
\label{sec-acceptance}
\begin{figure}[htpb!]
\centerline{\psfig{figure=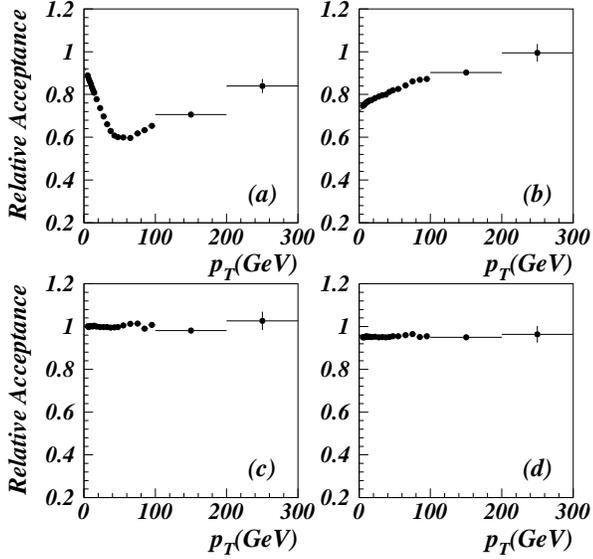,width=3.2in}}
\caption{Effect of requirements on (a) \et, (b) $\eta$,
(c) trigger \et\ cut-off, and (d) tracking efficiency
on the relative acceptance as a function of \pt.}
\label{FIGaccall}
\end{figure}

The parameterized detector simulation referred to in
Section~\ref{sec-data_selection_and_reconstruction} is used to
determine the overall acceptance as a function of \pt\ of the \zb\
boson. The effects of the trigger turn-on in \et, the rapidity
cut-offs, the $\phi$ module boundaries in the central calorimeter,
the pseudorapidity dependence of the tracking efficiency, and the
final \et\ requirements are all included in the calculation of the
acceptance. Figure~\ref{FIGaccall} shows the relative effects of
the requirements on the electron \et\ and pseudorapidity, and of the
trigger and tracking efficiency on the acceptance as a function of
\pt. As can be seen, the strongest effects come from the electron 
\et\ and pseudorapidity requirements. The dip in relative acceptance 
seen in  Fig.~\ref{FIGaccall}(a) for middle values of \pt\ results 
from one of the electrons carrying most of 
the \pt\ of the \zb\ boson--one electron can have a relatively large 
\et\ while the other has relatively small \et. However, as the \pt\ 
of the \zb\ boson increases beyond $45$ \gevc, this asymmetry is no 
longer allowed -- both electrons must have relatively large \et. 
The monotonic rise of the relative acceptance in 
Fig.~\ref{FIGaccall}(b) is due to the increasing ``centrality'' of the
event--as \pt\ increases, the rapidity of the \zb\ boson is 
closer to zero. As can be seen in Figs.~\ref{FIGaccall}(c) and (d), 
the imposition of the other selection criteria merely changes the 
normalization and does not affect the shape as a function of \pt.

\begin{figure}[htpb!]
  \centerline{\psfig{figure=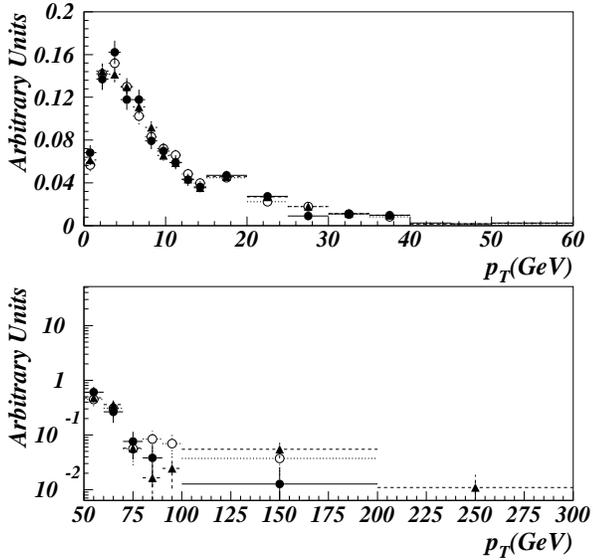,width=3.2in}}
  \caption{Comparison of the transverse momentum distribution
  of dielectron pairs with mass very close to the nominal \zb\
  boson mass, $90<M_{ee}<92$ \gevcc\ (solid circles) to those in
  the mass regions $75<M_{ee}<90$ \gevcc\ (open circles) and
  $92<M_{ee}<105$ \gevcc\ (closed triangles).}
  \label{FIGmasscomp}
\end{figure}

The mass requirement on the dielectron pairs has been ignored in
the final acceptance calculation. Figure~\ref{FIGmasscomp} compares
the \pt\ distribution for dielectron pairs with invariant mass near
that of the \zb\ boson to those with invariant mass above and below
the nominal \zb\ boson mass, and supports the expectation that any
\pt\ dependence on mass (near the \zb\ boson mass peak) is very
small.

\begin{figure}[htpb!]
\centerline{\psfig{figure=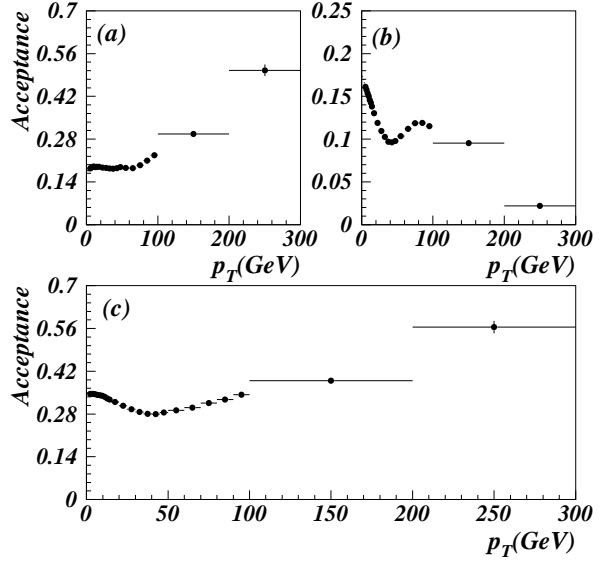,width=3.2in}}
\caption{Final acceptance as a function of \pt\ for (a) the
CCCC, (b) CCEC event topologies, and (c) for the
combined event sample.}
\label{FIGaccfinal}
\end{figure}

Figure~\ref{FIGaccfinal} shows the acceptance for the CCCC and CCEC
event topologies, as well as for the combined event sample. Here
we see the increased centrality of the events as a function of \pt,
noting the increasing acceptance for the CCCC events in contrast
to the decreasing acceptance for the CCEC events. The dip and rise
in Fig.~\ref{FIGaccfinal}(b) are due to competing effects of the
electron \et\ and pseudorapidity requirements.

\begin{figure}[htpb!]
\centerline{\psfig{figure=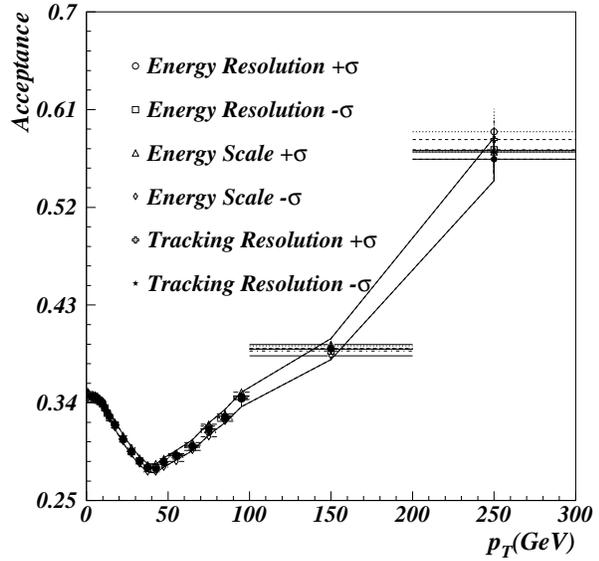,width=3.2in}}
\caption{Effect of uncertainties on the acceptance
in each bin of \pt. The band corresponds to a parameterization of
the uncertainty as a function of \pt, as discussed in the text.}
\label{FIGaccsys}
\end{figure}

The effect of uncertainties in the energy scale and resolution,
the tracking resolution, and the trigger efficiency is assessed for
each bin of \pt\ by varying the values of these parameters by their
measured uncertainties. Figure~\ref{FIGaccsys} shows the nominal acceptance
and those obtained by varying the values of the parameters. The
largest differences are observed at high \pt. If we parameterize
this systematic uncertainty as a linear function of
\pt, we obtain $\delta_{acc}=\pm (0.01+0.0001p_T)$. This
resulting band of uncertainty is also shown in
Fig.~\ref{FIGaccsys}.

\begin{figure}[htpb!]
  \centerline{\psfig{figure=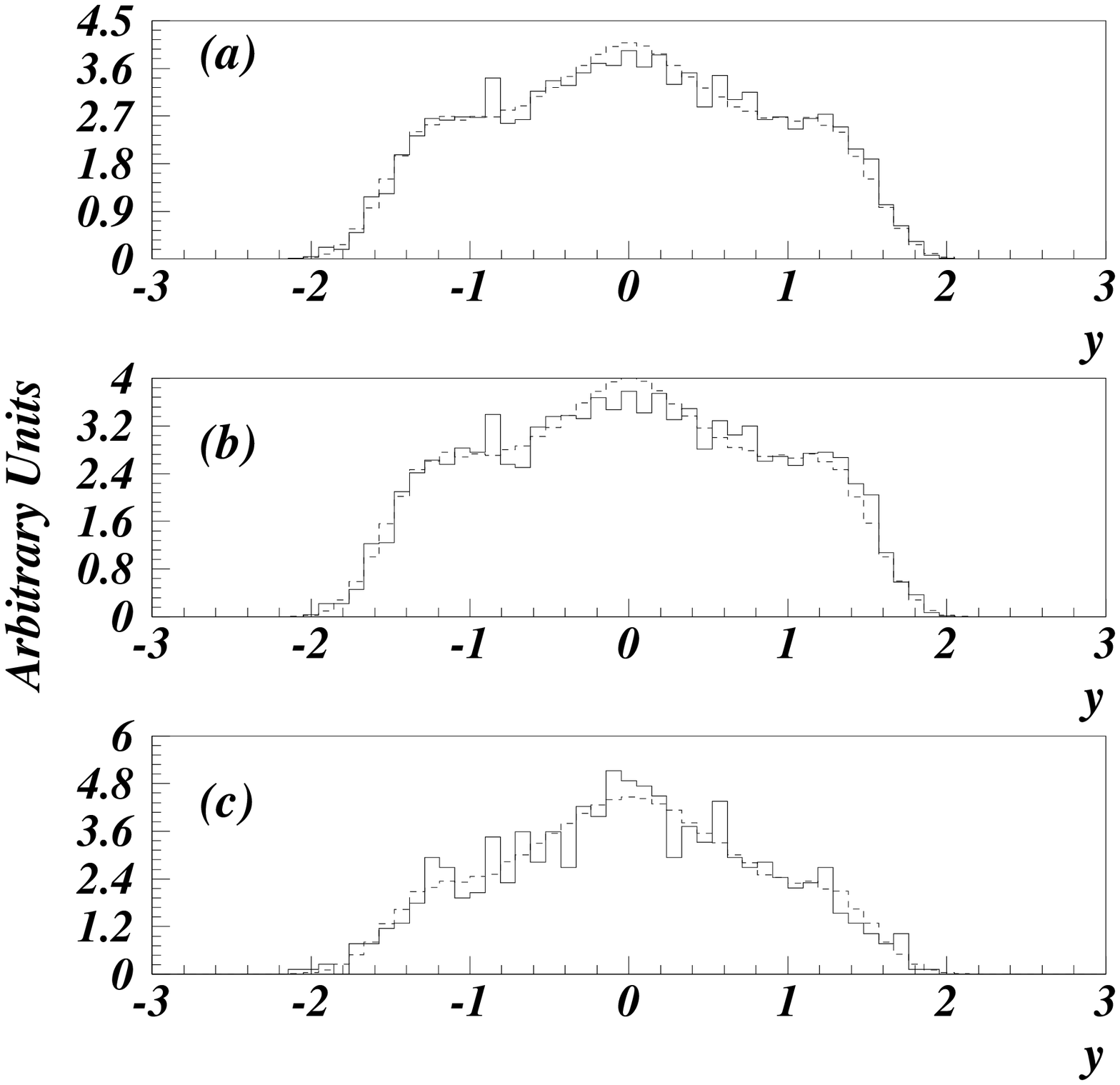,width=3.2in}}
  \caption{Comparison of \zb\ boson rapidity ($y$)
  distribution for the data (solid) and the simulation (dashed)
  for (a) all values of \pt, (b)
  \pt$<$20 \gevc, and (c) 20$<$\pt$<$50 \gevc.}
  \label{FIGdatamcrap}
\end{figure}

Because we determine the acceptance bin by bin in \pt, we are
relatively insensitive to the underlying model for the \pt\
spectrum used in the detector simulation. Nevertheless, we are
sensitive to the assumed rapidity distribution of the \zb\ boson
in each bin of \pt. The uncertainty in the predicted rapidity of the
\zb\ boson is expected to be dominated by the uncertainty in the pdf's
used for modeling \zb\ production. The uncertainty in acceptance
due to the choice of pdf
has been found to be $\pm 0.3\%$ for the inclusive measurement of
the \zb\ boson cross section\cite{wzcross}. This constrains the
uncertainty in the low-\pt\ region, where the cross section is
largest, to a value that is far smaller than the uncertainty from
variations in the parameters of the model of the detector.
Figure~\ref{FIGdatamcrap} shows that the rapidity distributions
obtained from the detector simulation and for data agree for both
low and high values of \pt; we therefore ignore any additional
uncertainty in the acceptance due to the modeling of the rapidity
of the \zb\ boson.

\section{Backgrounds}
\label{sec-backgrounds}
The primary background to dielectron production at the Tevatron is
from multiple-jet production from QCD processes in which the jets
have a large electromagnetic component (most of the energy is
deposited in the EM section of the calorimeter) or they are
mismeasured in some way that causes them to pass the electron
selection criteria. There are also contributions to the \zb\ boson
dielectron signal that are not from misidentification of electrons, but
correspond to other processes that differ from the one we are
trying to measure, e.g., \zbtt\ and \ttbar\ production. Such
processes are irreducible due to the fact that they have the same
final event signature as the signal, and often have
\pt\ dependences that can differ from the
\zb /$\gamma^*$ mediated production of the \zb\ boson and Drell-Yan
pairs. These must be determined and accounted for in any comparison
of data with theory.

Both the normalization and the shape of the multijet background as a
function of \pt\ are determined from data. Three types of
backgrounds have been studied to examine whether differences in
production mechanism or detector resolution would produce a
significant variation in the background: dijet events (from multijet
triggers), direct-$\gamma$ events (from single photon triggers),
and dielectron events in which both electrons failed the quality
criteria (from the \zb\ boson trigger). For the dijet events, we
selected the two highest-\et\ jets and reconstructed the ``\zb\
boson" as if the jets were electrons. Similarly, for the
direct-$\gamma$ events, we selected the highest-\et\ photon
candidate and the highest-\et\ jet in the event. For the
failed-dielectron sample, we used the two highest-\et\ electron
candidates whose cluster shape variable ($\chi_{HM}^2$) did not
match well with that of an electron. For all three backgrounds, the
``electron" objects were required to satisfy the same \et\ and
$\eta$ criteria as the data sample.

\begin{figure}[htpb!]
\centerline{\psfig{figure=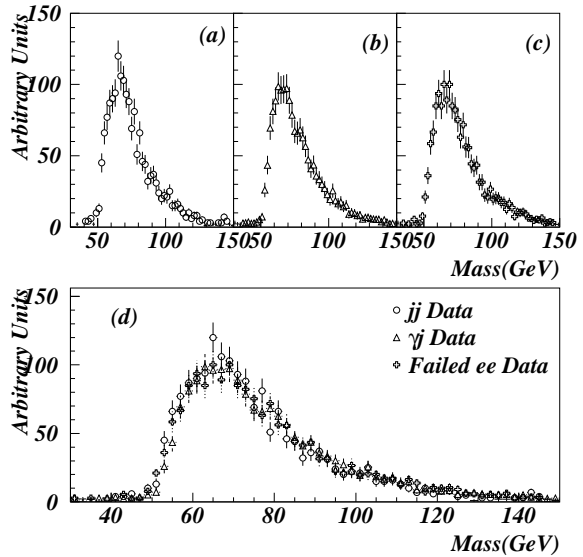,width=3.2in}}
\caption{Invariant mass distributions for the three types of
QCD multijet background samples in the CCCC topology: (a) dijet
data sample, (b) direct-$\gamma$ data sample, (c) failed-dielectron data 
sample. We show the distributions for all three data samples in (d).}
\label{fig:bakmass_cccc}
\end{figure}

\begin{figure}[htpb!]
\centerline{\psfig{figure=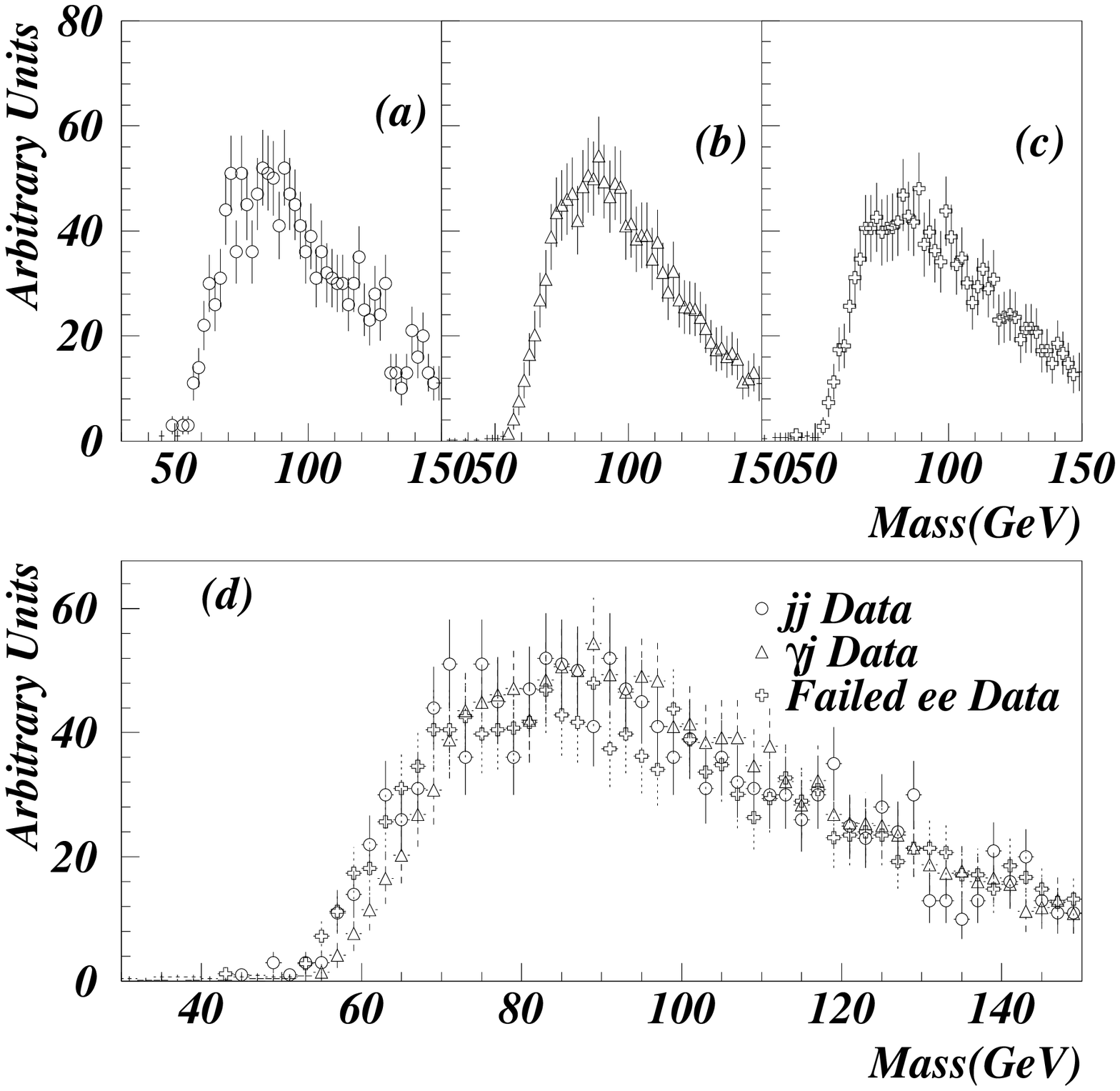,width=3.2in}}
\caption{Invariant mass distributions for the three types of
QCD multijet background samples in the CCEC topology: (a) dijet
data sample, (b) direct-$\gamma$ data sample, (c) failed-dielectron data 
sample. We show the distributions for all three data samples in (d).}
\label{fig:bakmass_ccec}
\end{figure}

\begin{figure}[htpb!]
\begin{tabular}{cc}
\centerline{\psfig{figure=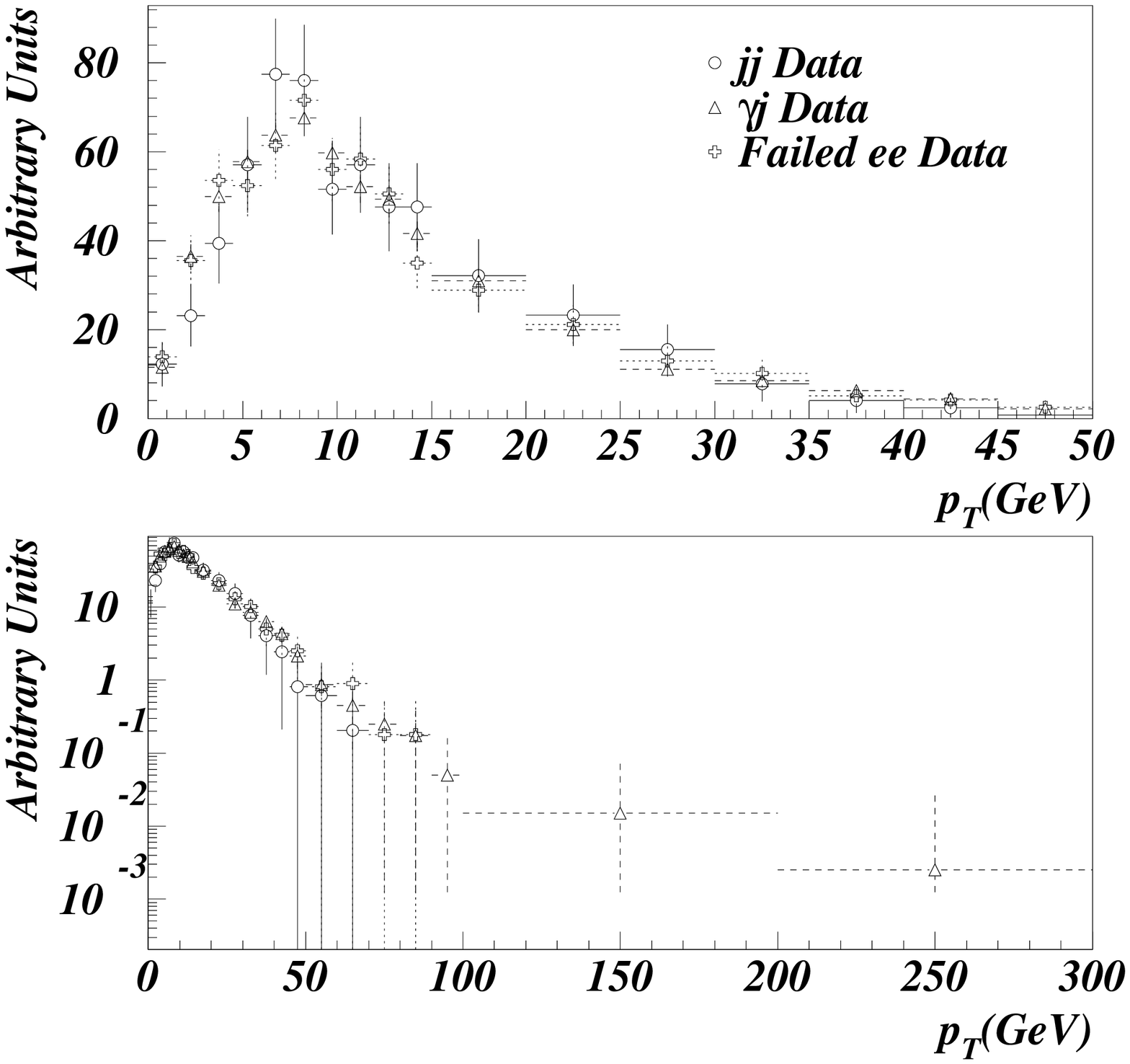,width=3.2in}}
\end{tabular}
\caption{Comparison of shapes of transverse momentum
distributions for the three multijet background samples for the
CCCC event topology.}
\label{fig:bakpt_cccc}
\end{figure}

\begin{figure}[htpb!]
\begin{tabular}{cc}
\centerline{\psfig{figure=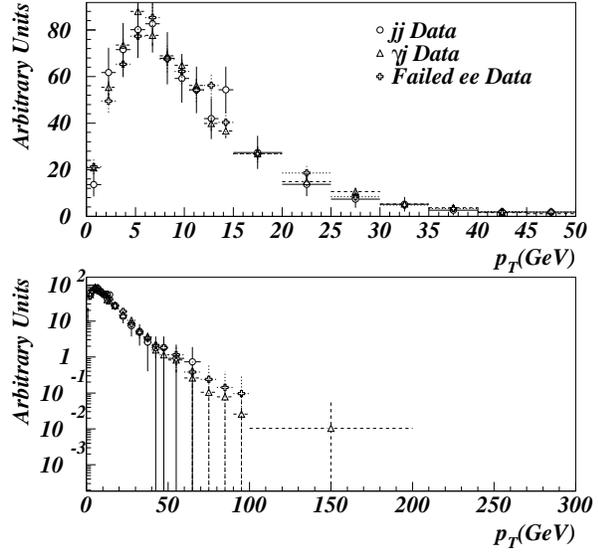,width=3.2in}}
\end{tabular}
\caption{Comparison of shapes of transverse momentum
distributions for the three multijet background samples for the
CCEC event topology.}
\label{fig:bakpt_ccec}
\end{figure}

Figures~\ref{fig:bakmass_cccc}--\ref{fig:bakpt_ccec} show the
invariant mass and \pt\ distributions for the background samples in
both the CCCC and CCEC event topologies. The direct-photon and
failed-dielectron events agree in the mass and \pt\ distributions.
The Kolmogorov-Smirnov probability ($P_{KS}$) for the two mass
distributions is 0.78 and between the two
\pt\ distributions it is 0.97. The
\pt\ distribution from the dijet sample also agrees well with the
direct-$\gamma$ and failed-dielectron samples, with $P_{KS}=0.51$
and $P_{KS}=0.57$, respectively. The dijet mass distribution does
not agree as well, giving $P_{KS}=0.005$ when comparing to the
direct-$\gamma$ sample and $P_{KS}=0.1$ when comparing to the
failed-dielectron sample. The difference is likely due to the
poorer jet-energy resolution compared to the electron energy
resolution. This difference in the shape of the invariant mass is
included in the systematic uncertainty on the background
normalization, and is a small effect (see
Section~\ref{backgroundlevel}).

\subsection{Multijet Background Level}
\label{backgroundlevel}
Because the mass distribution for the multijet background samples
depends on event topology, the level of the multijet background is
determined separately for CCCC and CCEC dielectron events. Using
this background, and the contribution from the \zb\ boson, we can
obtain the relative background fraction through a
maximum-likelihood fit for the amount of background and signal in
the data.

\begin{figure}[htpb!]
\centerline{\psfig{figure=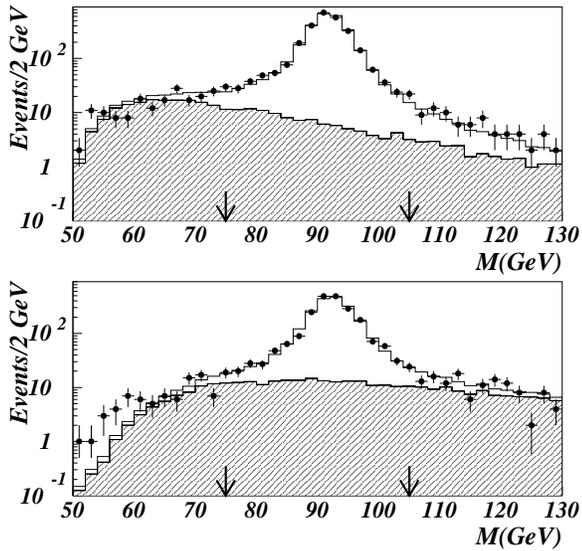,width=3.2in}}
\caption{Comparison of the dielectron invariant mass distribution
(closed circles) and the background
(dashed line) to the fit to \PYTH\ \zb/$\gamma^*$ and
background (solid line) for (a) CCCC and (b) CCEC topologies.}
\label{fig:backmassfit}
\end{figure}

We use the \PYTH\ event generator~\cite{PYTHIA} to produce the
invariant mass spectrum for the signal. Contributions from both \zb\
boson and Drell-Yan production and their quantum-mechanical interference
are included in the calculation. The generated four-momenta are
smeared using the detector simulation described previously. We
obtain the amount of multijet background in the data by performing
a binned maximum-likelihood fit to the sum of the signal (\PYTH)
and background:
\begin{equation}
N_{\rm data}(m_i)=c_1 N_{\PYTH}(m_i) + c_2 N_{\rm background}(m_i)
\end{equation}
where $c_1$ and $c_2$ are the normalization factors for the signal
and background contributions, respectively, and $m_i$ is the $i$th
mass bin. The fit was performed in the dielectron invariant mass
window of $60<M_{ee}<120$ \gevcc. Figure~\ref{fig:backmassfit}
shows the best fit to the dielectron invariant mass, separately for
CCCC and CCEC topologies using the direct-$\gamma$ sample as the
background. Using the other two background samples yields similar
results. The final value for the fraction of multijet background in
the data, $f_{\rm back}$, is defined by normalizing the fit parameter
$c_1$ to the number of events observed in the mass window of the \zb\
boson ($75<M_{ee}<105$ \gevcc):
\begin{equation}
f_{\rm back} = c_1 {N_{\rm total}({\rm data})\over
                    N_{\rm total}({\rm background})}
                    {N_{75-105}({\rm background})\over
                    N_{75-105}({\rm data})}
\end{equation}
where
\begin{equation}
N_{\rm total}({\rm sample}) = \sum_{\rm all\ m_i}
N_{\rm sample}(m_i)
\end{equation}
\begin{equation}
N_{75-105}({\rm sample}) = \sum_{ 75<m_i<105}
N_{\rm sample}(m_i).
\end{equation}

We use the direct-$\gamma$ sample for the central value of the
level of multijet background, and use the statistical uncertainty
from that fit. We also assign a systematic uncertainty associated
with our choice of mass window used in the fit and for differences
in the background models. We assign a systematic uncertainty to the
background normalization that corresponds to half of the maximum
difference from the central value in the determined background 
fractions. The background
values for each topology and the resulting uncertainties are
summarized in Table~\ref{tab:backfrac}.

Combining the uncertainties in quadrature, we obtain a background
fraction of ($2.45\pm0.50$)\% for the CCCC topology and
($7.09\pm1.00$)\% for the CCEC topology. Weighting the background
fractions by the relative number of events in each topology, we
obtain a total multijet background level of (4.44$\pm$0.89)\%.

\begin{table}
\begin{center}
\begin{tabular}{|c|c|c|}
\hline
Background Model   & CCCC             & CCEC \\
\hline
direct-$\gamma$    & ($2.45\pm0.41$\%)  & ($7.09\pm0.87$\%) \\
$55<M_{ee}<125$    & ($2.10\pm0.36$\%)  & ($7.52\pm0.83$\%) \\
$65<M_{ee}<115$    & ($2.74\pm0.51$\%)  & ($6.84\pm0.96$\%) \\
dijets             & ($1.98\pm0.35$\%)  & ($6.37\pm0.80$\%) \\
failed dielectrons & ($2.10\pm0.37$\%)  & ($6.22\pm0.78$\%) \\
\hline\hline
model uncertainty  & 0.24\%           & 0.44\% \\
window uncertainty & 0.17\%           & 0.22\% \\
\hline
\end{tabular}
\caption{Background fractions in the two primary event topologies. 
The values in the first five rows include only statistical 
uncertainties for each method. The
systematic uncertainties obtained by considering variations in the 
background selection and fitting criteria are shown in the last two 
rows.}
\label{tab:backfrac}
\end{center}
\end{table}

\subsection{\pt-Dependence of the Multijet Background}

The direct-$\gamma$ sample is used to determine the shape of the
background distribution for several reasons. First, this sample has
the greatest number of events. Second, we expect the
direct-$\gamma$ data sample to provide a good approximation of the
combination of backgrounds from dijet and true direct-$\gamma$
production because about half of the
direct-$\gamma$ sample consists of misidentified dijets, and
therefore has the approximate balance of dijet and direct-$\gamma$
events expected from QCD sources. Third, since events in the direct-$\gamma$
often contain at least one good electron-like object, detailed
differences between choosing electron-like objects and jet objects
for reconstructing the ``\zb " boson are smaller here.

The final shape of the background is obtained by combining the CCCC
and CCEC samples, weighted by the relative contributions to the
background. To facilitate later analysis, the shape is
parameterized as a function of \pt\ using the following functional
forms:
\begin{equation}
  \begin{array}{cc}
   a(p_T+b)^2 e^{\alpha p_T}&
                   {\rm if}\ (p_T< 8\ {\rm GeV})  \\
   a({1\over p_T})^2+be^{\alpha p_T}&
                   {\rm if}\ (p_T> 8\ {\rm GeV}) \\
   \end{array}
\end{equation}

\begin{figure}[htpb!]
\centerline{\psfig{figure=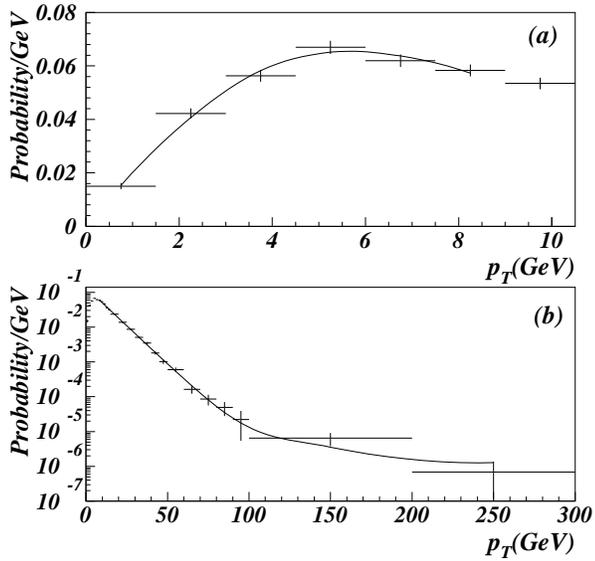,width=3.2in}}
\caption{The results of the fit to the background as
a function of \pt. For (a) \pt$<$8 \gevc, the
$\chi^2/{\rm d.o.f.}=0.9$, and (b) for \pt$>$8 \gevc,
the $\chi^2/{\rm d.o.f.}=0.7$.}
\label{fig:backfit}
\end{figure}

The function is normalized to be a probability distribution, that is
the product of the function and the total number of background events
results in the differential background in each bin of \pt. 
Figure~\ref{fig:backfit} shows the results of the fit to the
background and Table~\ref{tab:backfitparam} shows the values of the
fit parameters. 

\begin{table}
\begin{center}
\begin{tabular}{|c|c|c|}
\hline
Parameter    &    $p_T<8$ \gevc  & $p_T>8$ \gevc  \\
\hline\hline
$\alpha$   &    $-0.31\pm0.02$        & $-0.1014\pm0.0015$        \\
$a$        &    $0.001\pm0.002$      & $0.08\pm0.03$     \\
$b$        &    $0.65\pm0.18$         & $0.136\pm0.004$ \\
\hline
\end{tabular}
\caption{Values of the parameters obtained from the fit to the
direct-$\gamma$ background.}
\label{tab:backfitparam}
\end{center}
\end{table}

\subsection{Other Sources of Dielectron Signal}

Although \zbee\ and QCD multijet events make up nearly all of
observed dielectron signal, there 

are contributions from other
sources, such as \zbtt, \ttbar, and diboson ($WW$, $ZZ$, $WZ$,
$W\gamma$, $Z\gamma$)
production in dielectron final states. The expected contributions
from these sources are estimated below.

The dielectron event rate from  $Z\rightarrow\tau^+ \tau^-$ production
in our accepted mass range is calculated to be $<2.6\times10^{-6}$
per \zbee\ event\cite{Jiang_thesis}. The events were generated with
the \HERW\ simulator and smeared with the \Dzero\ detector resolutions.
For the current sample, this
corresponds to less than $0.009$ events for all values of
dielectron \pt. We therefore ignore this contribution to the signal.

The dielectron background contribution from \ttbar\ production is
concentrated at high \pt. The fraction was determined using
the \HERW\ simulator for \ttbar\ production, smeared with the known
\Dzero\ detector resolutions. Electron
contributions from both $W\rightarrow e\nu$ and $W\rightarrow \tau
X \rightarrow eX$ channels were considered. For a
\ttbar\ cross section of 6.4 pb\cite{D0top}, and the standard
branching ratios for the \wb\ boson, the calculated geometric and
kinematic acceptance from \HERW\ is $0.01\pm0.006$. Including
electron identification efficiency for dielectron events, we expect
about $0.36$ events in the entire sample and about $0.2$ events
with dielectron \pt$> 50$ \gevc. Considering the small number of
events expected, the \ttbar\ contribution is also ignored.

\begin{table}
\begin{center}
\begin{tabular}{|c|p{2.5cm}|p{2cm}|p{1.5cm}|}
\hline
Process    &  Acceptance & $\sigma \cdot B$ (pb) & Expected events
\\
\hline\hline
\ttbar     &    0.01$\pm$0.006 &  0.43$\pm$0.01   &  $\approx$0.2 \\
$W\gamma$  &     $<$0.0003     &  11.3$\pm$0.3    &  $<.3$ \\
$WW$       &   0.016$\pm$0.007 &  0.12$\pm$0.03   &  $\approx$0.15 \\
$WZ$       &   0.016$\pm$0.007 &  0.08$\pm$0.01   &  $\approx$0.1 \\
$ZZ$       &   0.046$\pm$0.002 &  0.03$\pm$0.01   &  $\approx$0.005\\
\hline
\end{tabular}
\caption{The expected number of events from diboson and top quark processes.
Each channel assumes a total luminosity of 108.5 \ipb, and
a dielectron identification efficiency of 0.73.}
\label{tab:rareback}
\end{center}
\end{table}

We considered $WW$, $ZZ$, $WZ$,  and $W\gamma$ events generated with
the \HERW\ simulator and smeared with the known \Dzero\ detector
resolutions. All of these backgrounds are small, and we therefore
focus on any possible effects on our measurement at high-\pt, where
there are relatively few events and effects of even a small
background contamination could be significant.

The resulting acceptances and expected number of background events
with \pt$>$50 \gevc\ are given in Table~\ref{tab:rareback}. No
$W\gamma$ events out of approximately 3000 generated passed the
selection requirements, because very few such events have photons
with \et$>$25 \gev\ and an invariant mass ($M_{e\gamma}$) near the
\zb\ boson mass. The table includes the assumed production cross sections
multiplied by branching ratios ($\sigma\cdot B$) for
\wb\ and \zb\ boson into electron states. The $WW$ cross section
($10.2^{+6.3}_{-5.1}$ pb) and branching ratio to dielectrons
(0.011) are obtained from Ref.~\cite{wwcross}. The value
$\sigma(W\gamma)\cdot B(W\rightarrow e\nu) = 11.3^{+1.7}_{-1.5}$ pb, is
obtained from Ref.~\cite{wgamma}, and assumes $p^{\gamma}_T>10$
\gevc\ and $\Delta R_{e\gamma}>0.7$. The standard model $WZ$
cross section is taken from Ref.~\cite{zephagwood}, and
the $ZZ$ cross section is taken from Ref.~\cite{eitchen}. Given their
small size, all of these contributions have been ignored in our
analysis.

\section{Measured \dsdpt}
\label{sec-smearxs}
Table~\ref{tab:prd_results} shows the values for each of the
individual components of the measurement: the number of events
observed for each bin of \pt, the product of the efficiency and the
acceptance ($\epsilon\cdot a$), and the expected number of background
events ($b$). The associated uncertainties are also included. We
combine the geometric acceptance and identification efficiencies into
a single overall event efficiency by taking their product. We assume
that the uncertainties are well-described as Gaussian distributions
and add them in quadrature to obtain the uncertainty
$\delta(\epsilon\cdot a)$.

The measured differential cross section, \smdsdpt, is obtained by
calculating the cross section in each bin of \pt, accounting for
the effects of efficiency, acceptance, and background, but not
accounting for the effects of detector smearing. That is,
\begin{equation}\label{EQresult}
  ({d\sigma^\prime \over dp_T})_i =
  {\sigma^\prime_i\over\Delta_i^{\text bin}}.
\end{equation}
where $\sigma^\prime_i$ is the measured cross section in bin $i$ and
$\Delta_i^{\text bin}$ is the width of the bin in \pt.

We obtain the cross section and uncertainty in each bin using the
methods of statistical inference. We
relate the expected number of events $\mu$ in each bin to the
underlying cross section~\cite{Loredo,Jaynes}:
\begin{equation}\label{EQ_expevt}
  \mu={\cal{L}}\epsilon\sigma^\prime+b,
\end{equation}
where \lum\ is the total integrated luminosity, $\epsilon$ is the
overall detection efficiency for the process, and $b$ is the number
of background events. A value of $\mu$ is determined for each bin 
of \pt.

We relate the observed number of events and the expected number of
events through a probability distribution, in our case an assumed
Poisson distribution,
\begin{equation}\label{EQ_poisson}
  P(d|\sigma^\prime,\epsilon,b,{\cal{L}},I) =
  {e^{-({\cal{L}}\epsilon\sigma^\prime+b)}({\cal{L}}\epsilon\sigma^\prime+b)^d\over d!}.
\end{equation}
where $d$ is the number of events observed and $I$ refers to the
assumptions implicit in deriving the probability
density\cite{Jaynes}.

Applying Bayes' Theorem, we invert the probability in
Eq.~\ref{EQ_poisson},
\begin{multline}\label{EQ_bayes1}
  P(\sigma^\prime,\epsilon,b,{\cal{L}}| d,I) = \\
{P(d|\sigma^\prime,\epsilon,b,{\cal{L}},I)
  P(\sigma^\prime,\epsilon,b,{\cal{L}}|I)\over
  \zeta}
\end{multline}
where $\zeta$ normalizes the probability such that
$\int P(\sigma^\prime,\epsilon,b,{\cal{L}}|d,I)d\Omega \equiv 1$, where
$d\Omega$ denotes that the integration is over all relevant variables.
$P(\sigma^\prime,\epsilon,b,{\cal{L}}|d,I)$
is the joint posterior probability, describing the probability of a
particular set of $\sigma^\prime,\epsilon,b$, and \lum\, given the results
from our data. $P(\sigma^\prime,\epsilon,b,{\cal{L}}|I)$ is the joint
prior probability, describing the probability of a particular set
of $\sigma^\prime,\epsilon,b$, \lum\ before taking our data into account.
$P(d|\sigma^\prime,\epsilon,b,{\cal{L}},I)$ is the likelihood function for
our data. Assuming that the individual parameters are logically
independent, \eg, the cross section does not depend on the
background, then Eq.~\ref{EQ_bayes1} can be rewritten as
\begin{multline}\label{EQ_bayes2}
  P(\sigma^\prime,\epsilon,b,{\cal{L}}| d,I) = \\
  {P(d|\sigma^\prime,\epsilon,b,{\cal{L}},I)
  P(\sigma^\prime|I)P(\epsilon|I)P(b|I)P({\cal{L}}|I)\over
  \zeta}.
\end{multline}

We are not interested in the values of the parameters $\epsilon$,
$b$, and \lum, and we eliminate the dependence of the joint
posterior probability on these nuisance parameters by integrating
over their allowed values, 

\begin{figure}[htpb!]
  \centerline{\psfig{figure=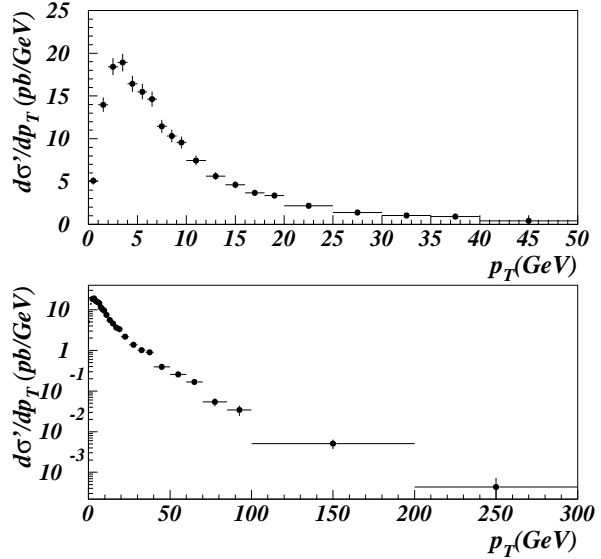,width=3.2in}}
  \caption{The measured
  differential cross section, not corrected for detector smearing,
   (a) for \pt$<$50 \gevc\ and (b) for all \pt. }
  \label{fig:meas_results}
\end{figure}

a process called marginalization. To
extract our results, we calculate
\begin{multline}\label{EQ_psig}
  P(\sigma^\prime|d,I) = 
  \int dbd{\cal{L}} d\epsilon d\sigma^\prime  \\
  { P(d|\sigma^\prime,\epsilon,b,{\cal{L}},I)
  P(\sigma^\prime|I)P(\epsilon|I)P(b|I)P({\cal{L}}|I)\over
  \zeta} .
\end{multline}
In the calculation of the binned differential cross section, the
uncertainty on the integrated luminosity changes only the overall
normalization of the distribution, which is already accounted for in
our normalization to the \Dzero\ measurement of $\sigma_{Z\rightarrow
ee}$. We therefore use a delta function as the prior for the
integrated luminosity distribution. We assume the priors for the
efficiency ($\epsilon$) and background ($b$) to be Gaussian
distributed, with their estimated mean values and standard deviations
as the means and widths of the Gaussians. The prior probability
distribution for the cross section in each bin is taken to be
independent of $\sigma$ (uniform for the range
$[\sigma_{min},\sigma_{max}]$ where $\sigma_{min}>0$) and the total
range is at least $\pm$6 standard deviations around the mean.

\begin{figure}[htpb!]
  \begin{tabular}{l l}
  \psfig{figure=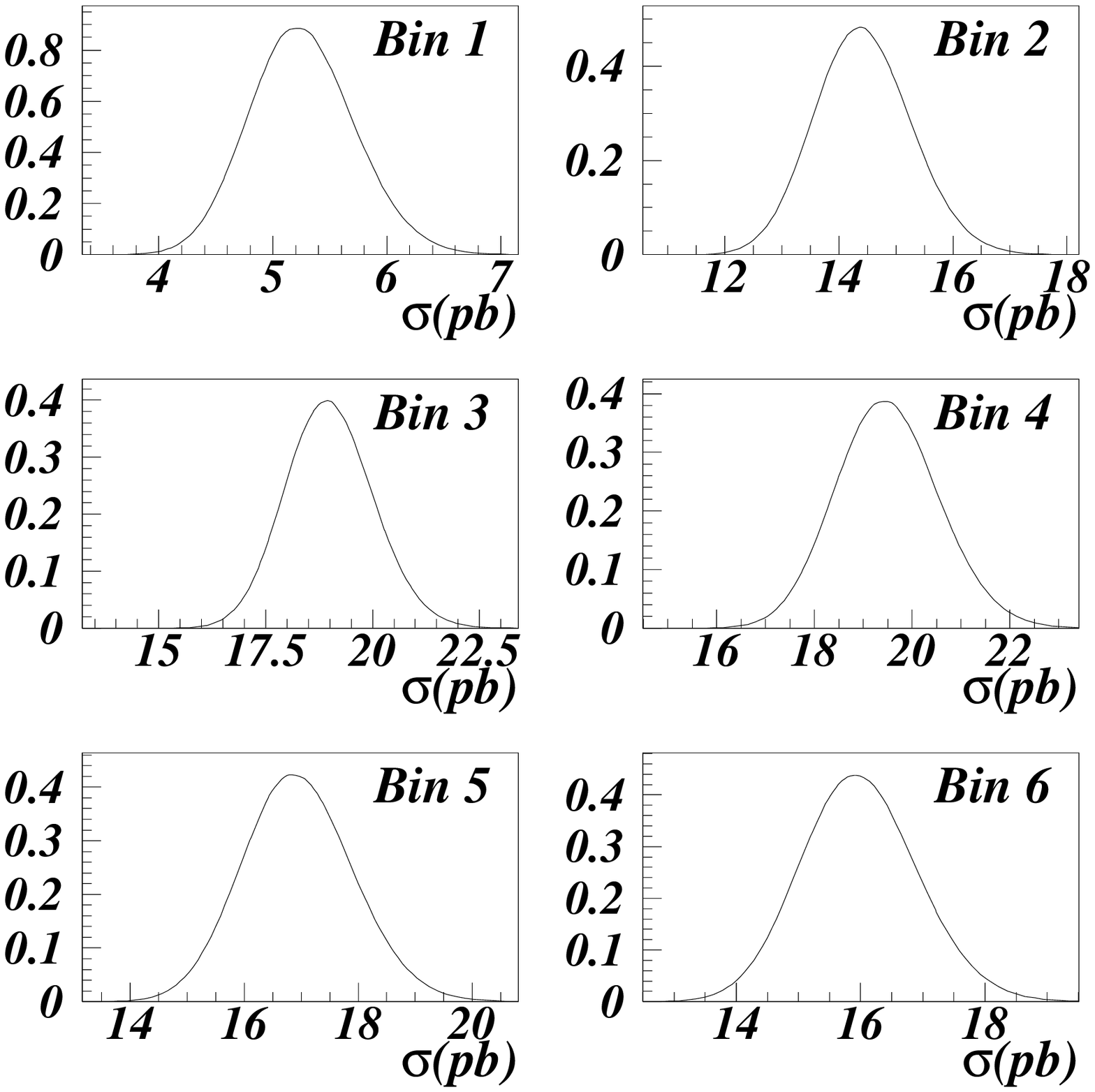,width=3.2in} &
  \psfig{figure=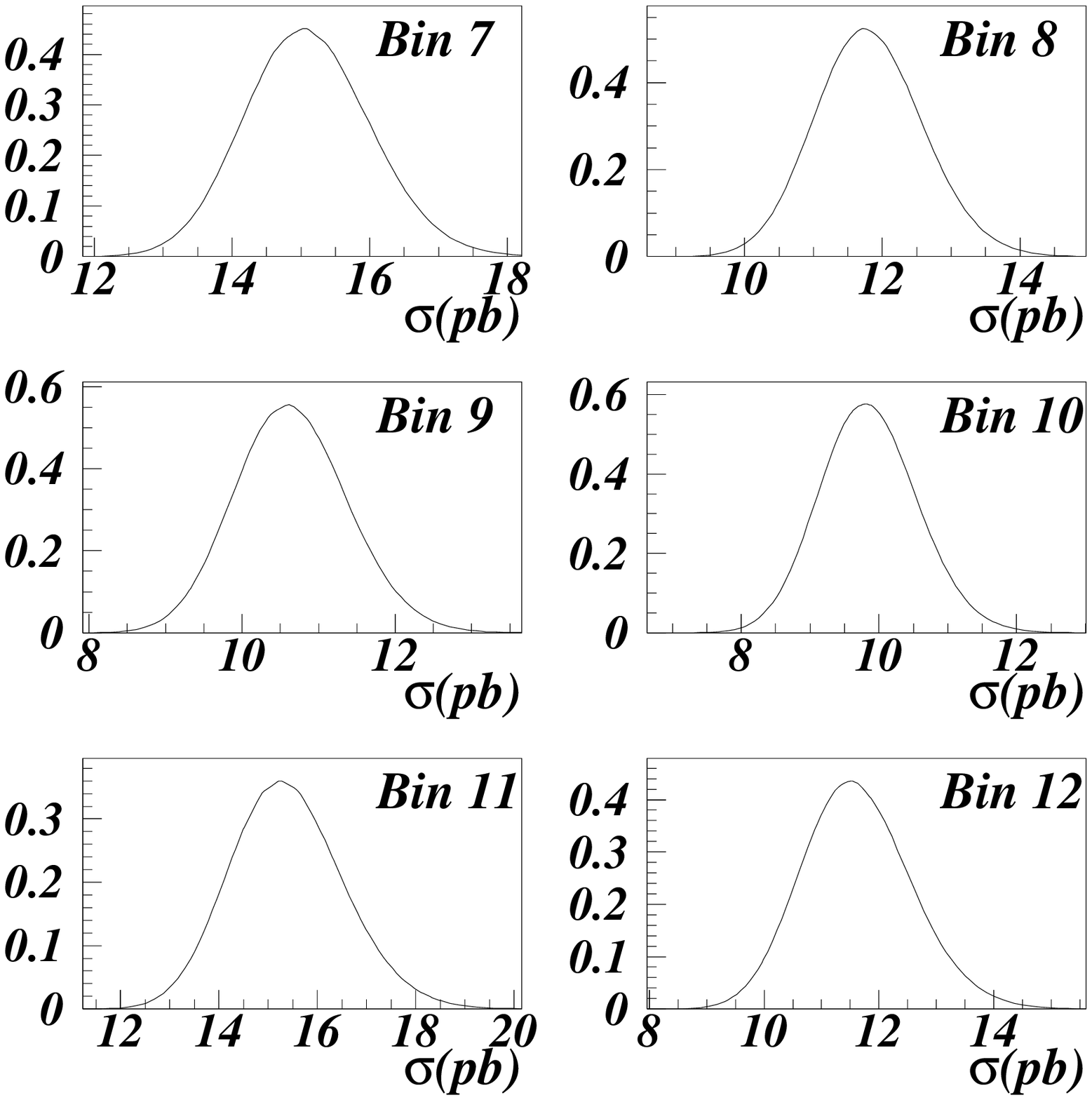,width=3.2in} \\
  \psfig{figure=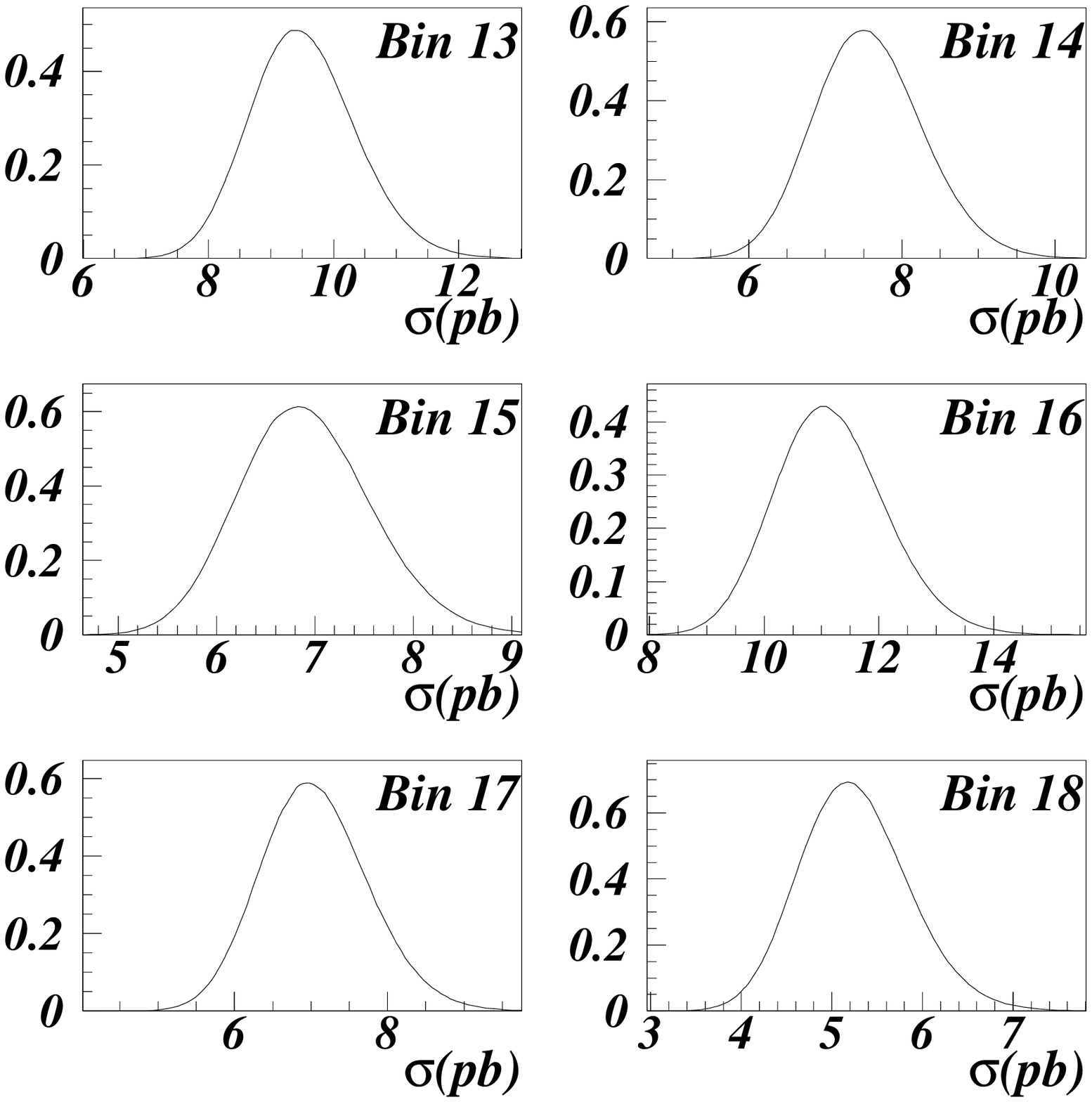,width=3.2in} &
  \psfig{figure=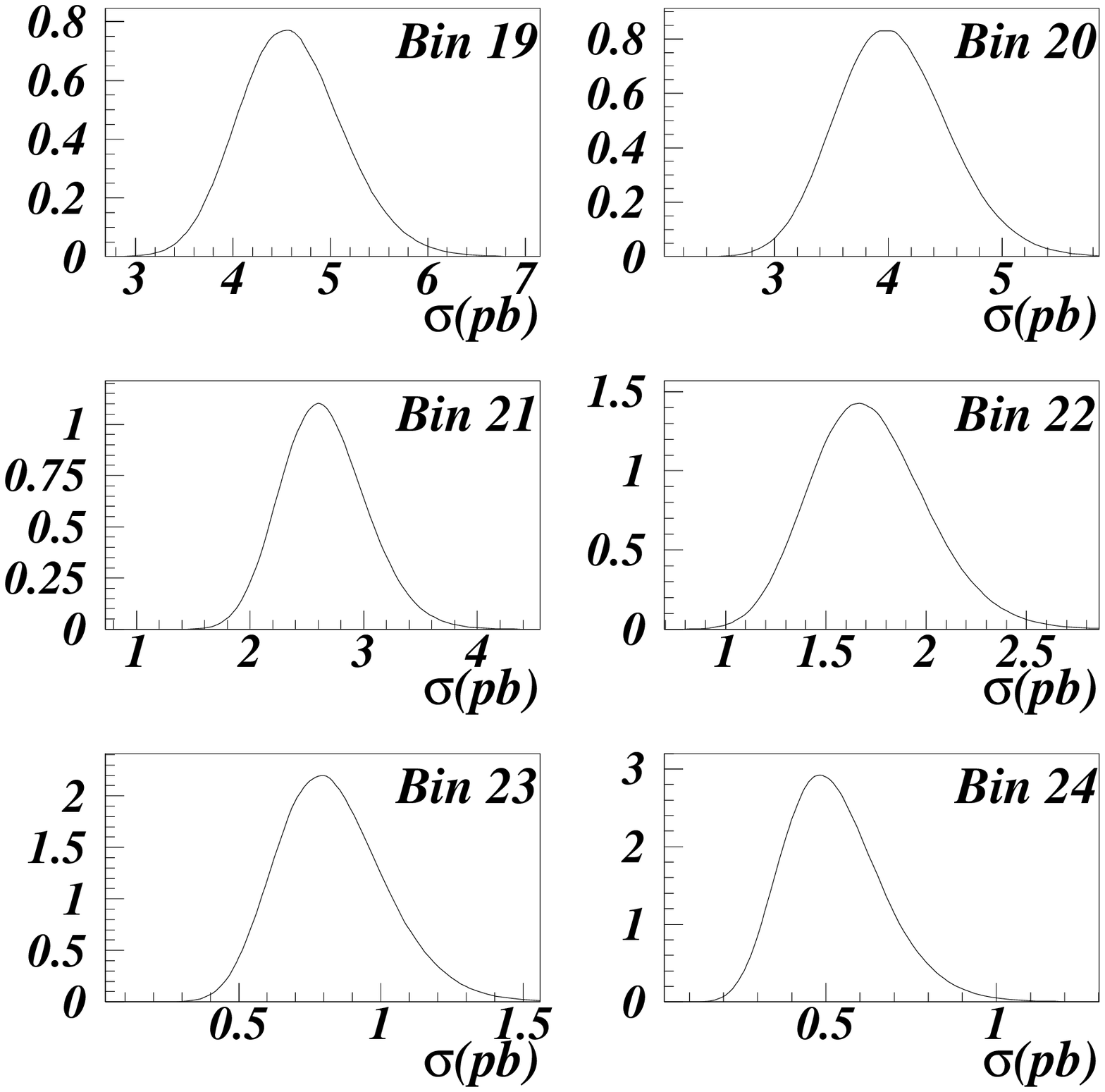,width=3.2in} \\
  \end{tabular}
  \centerline{\psfig{figure=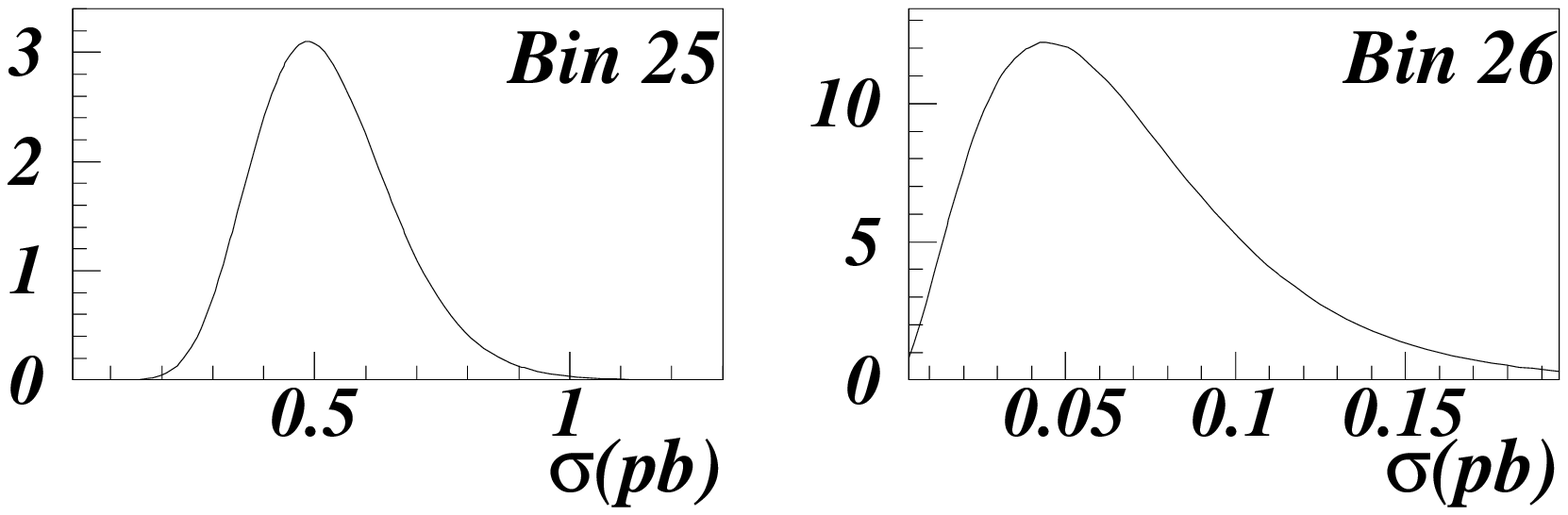,width=3.2in}} \\
  \vspace{-5.0cm}
  \caption{Normalized probability distributions for
  cross sections ($\sigma$) in individual bins of \pt. The vertical
scale in each case is Probability/pb.}
   \label{fig:xsect}
\end{figure}

The integration in Eq.~\ref{EQ_psig} is performed using the
numerical integrator \MISE~\cite{MISER}, and the results are shown
in Fig.~\ref{fig:xsect}. Since the probability distributions for
all but the highest-\pt\ bin are nearly Gaussian, we assign the 
final value of the cross section for each bin in \pt\ to be the 
mean of the probability distribution with uncertainties set equal 
to the standard deviation about the mean. For the highest-\pt\ bin, we 
use the most probable value for the cross section with upper and lower
uncertainty values circumscribing the narrowest 68\% confidence interval.
The integral over \pt\ of the differential distribution is
normalized to the inclusive cross section for \zb\ boson production
measured by \Dzero\ \cite{wzcross}. 
Table~\ref{tab:prd_results} gives the values of
the measured differential cross section in each bin of \pt, not corrected for 
detector smearing, and Figure~\ref{fig:meas_results} displays the results 
as a function of \pt.

\section{Fit to Non-perturbative Parameters}
\label{sec-fitgvalues}
As discussed in Section~\ref{sec-introduction}, the current
theoretical understanding of the \pt\ distribution of \zb\ bosons
uses fixed-order perturbative calcuations to describe the high-\pt\
region and resummation calculations of the perturbative solution to
describe the low-\pt\ region. At the smallest values of \pt, a
parameterization must be invoked to account for
non-perturbative effects that are not calculable in perturbative
QCD. The generic form for the function is given in Eq.~\ref{eq:snpgen},
however, one must choose particular functional forms for $h_1(x,b)$ and
$h_2(b)$. Historically there are two versions for the choice of
this parameterization. The first, from Davies, Weber, and
Stirling (DWS)\cite{DWS}, has the form:
\begin{equation}\label{eq:dws}
  S_{NP}^{DWS}(b,Q^2)= g_1b^2+g_2b^2\ln({Q^2\over Q_o^2}).
\end{equation}
The values of $g_1$ and $g_2$ are determined by fitting to
low-energy Drell-Yan data, yielding $g_1=0.15$ \gevsq\ and $g_2=0.4$
\gevsq, where $Q_o=2$ \gev\ and $b_{max}=0.5$ \gev$^{-1}$ (see
Eq.~\ref{eq:wbstar}). They used the pdf's of Duke and
Owens~\cite{dukeowens}. The second is from Ladinsky and
Yuan\cite{LadinskyYuan}:
\begin{eqnarray}\label{eq:ly}
  S_{NP}^{LY}(b,Q^2)= \nonumber \hspace{-1.5cm} \\
  & g_1b^2+g_2b^2\ln({Q^2\over Q_o^2})+
                      g_1g_3b\ln(100x_ix_j)
\end{eqnarray}
where $x_i$ and $x_j$ are the momentum fractions of the colliding
partons. The values of $g_1$, $g_2$, and $g_3$ are determined by
fitting to low-energy Drell-Yan data and a small sample of \zbee\ data
from the 1988--89 run at CDF\cite{CDF_ptz}, yielding
$g_1=0.11^{+0.04}_{-0.03}$
\gevsq, $g_2=0.58_{-0.2}^{+0.1}$
\gevsq, and $g_3=-1.5_{-0.1}^{+0.1}$ \gev$^{-1}$, where
$b_{max}=0.5$ \gev$^{-1}$ and $Q_o=1.6$ \gev. They used the CTEQ2M 
pdf's~\cite{cteq2m} in the fits.

The \zb\ boson \pt\ distribution is by far most sensitive to the value
of $g_2$. For measurements at the Tevatron at $Q^2=M_Z^2$, the
calculation is nearly insensitive to the value of $g_3$, and only
slightly sensitive to the value of $g_1$. For $g_3$, this
insensitivity is due to the high energy of the \ppbar\ beam relative
to the $Q^2$ being probed. For a center-of-mass energy of
$\hat{s}=x_ix_js$ and $\hat{s}\approx M_Z^2$, we see that for a
measurement at the Tevatron ($\sqrt{s}=1.8$ TeV), the $g_3$ term
becomes $g_1g_3b\ln(100\hat{s}/s) \approx -1.4g_1g_3b$. The $g_2$ term
varies as $g_2b^2\ln(M_Z/Q_o)\approx 4.7g_2b^2$, and therefore makes a far
larger contribution to the value of $S_{NP}$. The relative importance
of $g_2$ over $g_1$ comes from the $\ln({Q^2/Q_o^2})$ term.

Because the width of the \zb\ boson is $\approx2.5$ \gevcc, for purely
phenomenological needs the non-perturbative physics can be
parameterized using a single parameter $g^\prime = g_1 +
g_2\log({M_Z^2/Q_o^2})$~\cite{ellisVeseli}. However, because the general 
form of $S_{NP}$ is theoretically motivated, we preserve the form of
Eq.~\ref{eq:ly}, focusing on the value of $g_2$, the parameter we
are most sensitive to.

We perform a minimum-$\chi^2$ fit to determine the best
value of $g_2$ from our data. For the purposes of the fit, we fix
$g_1=0.11$ \gevsq\ and $g_3=-1.5$ \gev$^{-1}$, as suggested by
Ladinsky and Yuan~\cite{LadinskyYuan}. We use the program
\LEGA~\cite{BalazsYuan} with the CTEQ4M pdf's~\cite{cteq4m}
to generate the $d\sigma^\prime / dp_T$ distribution for the \zb\ boson
and match the low-\pt\ and high-\pt\ regions using the prescription in
\RESBO, obtaining a single grid for all values of \pt\ calculated to
NNLO. We smear the prediction with the \Dzero\ detector resolutions
and fit the resulting \pt\ distribution to our measured result.
The $\chi^2$ distribution as a function of $g_2$ is well-behaved and 
parabolic and when fit to a quadratic function
yields a value of 0.59$\pm$0.06 \gevsq\ at the minimum, with  
$\chi^2$/d.o.f$=10.6/10$.

For completeness, we also fit for the individual values of $g_1$ and
$g_3$, using the Ladinsky and Yuan values for the two parameters not
being fitted. The variation in $\chi^2$ of $g_1$ and $g_3$ are also
well-behaved and parabolic, and the fit yields $g_1=0.09\pm 0.03$ 
\gevsq\ and $g_3=-1.1\pm0.6$ \gev$^{-1}$. 
The value of $g_1$ agrees with the Ladinsky-Yuan result, and
is of comparable precision. The value of $g_3$ also agrees with
the Ladinsky-Yuan result, but is far less precise.

\onecolumn
\begin{table}
\begin{center}
\begin{tabular}{|c|c|c|c|c|c|c|c|c|c|c|c|}
  \hline
   Bin  & \pt\ range & number &  &
  & $b$ & $\delta b$
  & ${d\sigma^\prime/ dp_T}$ & $\delta({d\sigma^\prime/dp_T})$
  &  & ${d\sigma/dp_T}$ & $\delta({d\sigma/dp_T})$ \\
   number & (\gevc) & of events & $\epsilon\cdot a$ & $\delta(\epsilon\cdot a)$
  & (events) & (events)
  & (nb/\gevc) & (nb/\gevc)
  & $\alpha(p_T)$ & (nb/\gevc) & (nb/\gevc) \\
  \hline\hline
1  & 0--1    & 156 & 0.351 & 0.011 & 3.28  &  0.7   & 5.10    & 0.45   & 1.185  & 6.04    & 0.53 \\
2  & 1--2    & 424 & 0.347 & 0.011 & 8.14  &  1.6   & 14.0    & 0.82   & 1.160  & 16.2    & 0.96 \\
3  & 2--3    & 559 & 0.346 & 0.011 & 12.7  &  2.5   & 18.4    & 0.99   & 1.108  & 20.4    & 1.1  \\
4  & 3--4    & 572 & 0.343 & 0.011 & 16.1  &  3.2   & 18.9    & 1.0    & 1.042  & 19.7    & 1.1  \\
5  & 4--5    & 501 & 0.343 & 0.011 & 18.0  &  3.6   & 16.4    & 0.93   & 0.988  & 16.2    & 0.92 \\
6  & 5--6    & 473 & 0.342 & 0.011 & 18.8  &  3.8   & 15.5    & 0.90   & 0.965  & 15.0    & 0.87 \\
7  & 6--7    & 440 & 0.336 & 0.011 & 18.5  &  3.7   & 14.6    & 0.88   & 0.960  & 14.1    & 0.84 \\
8  & 7--8    & 346 & 0.335 & 0.011 & 17.5  &  3.5   & 11.5    & 0.76   & 0.967  & 11.1    & 0.73 \\
9  & 8--9    & 312 & 0.334 & 0.011 & 16.9  &  3.4   & 10.3    & 0.71   & 0.972  & 10.0    & 0.69 \\
10 & 9--10   & 285 & 0.330 & 0.011 & 15.2  &  3.1   & 9.55    & 0.69   & 0.972  & 9.29    & 0.67 \\
11 & 10--12  & 439 & 0.324 & 0.017 & 26.1  &  5.2   & 7.46    & 0.56   & 0.972  & 7.25    & 0.54 \\
12 & 12--14  & 326 & 0.317 & 0.017 & 21.3  &  4.3   & 5.63    & 0.46   & 0.967  & 5.45    & 0.44 \\
13 & 14--16  & 258 & 0.306 & 0.017 & 17.4  &  3.5   & 4.61    & 0.41   & 0.964  & 4.45    & 0.39 \\
14 & 16--18  & 203 & 0.302 & 0.016 & 14.2  &  2.8   & 3.67    & 0.34   & 0.963  & 3.54    & 0.33 \\
15 & 18--20  & 181 & 0.297 & 0.016 & 11.6  &  2.3   & 3.35    & 0.32   & 0.958  & 3.21    & 0.31 \\
16 & 20--25  & 287 & 0.289 & 0.016 & 20.5  &  4.1   & 2.16    & 0.19   & 0.954  & 2.06    & 0.18 \\
17 & 25--30  & 174 & 0.278 & 0.015 & 12.3  &  2.5   & 1.37    & 0.14   & 0.945  & 1.29    & 0.13 \\
18 & 30--35  & 124 & 0.270 & 0.016 & 7.46  &  1.5   & 1.02    & 0.12   & 0.944  & 0.962   & 0.11 \\
19 & 35--40  & 104 & 0.263 & 0.014 & 4.51  & 0.90   & 0.892   & 0.10   & 0.941  & 0.840   & 0.10 \\
20 & 40--50  &  92 & 0.264 & 0.014 & 4.38  & 0.88   & 0.392   & 0.048  & 0.952  & 0.373   & 0.045 \\
21 & 50--60  &  61 & 0.274 & 0.015 & 1.63  & 0.33   & 0.258   & 0.037  & 0.974  & 0.251   & 0.036 \\
22 & 60--70  &  40 & 0.283 & 0.016 & 0.616 & 0.12   & 0.167   & 0.028  & 0.975  & 0.163    & 0.027 \\
23 & 70--85  &  20 & 0.300 & 0.017 & 0.308 & 0.062  & 0.054   & 0.016  & 0.989  & 0.053   & 0.012 \\
24 & 85--100 &  13 & 0.319 & 0.018 & 0.095 & 0.019  & 0.034   & 0.010  & 0.988  & 0.034   & 0.009 \\
25 & 100--200 & 15 & 0.366 & 0.022 & 0.130 & 0.026  & 0.0051  & 0.0013 & 0.994  & 0.0050  & 0.0013 \\
26 & 200--300 &  2 & 0.530 & 0.034 & 0.038 & 0.0076 & 0.0004  & $^{+0.00038}_{-0.00029}$ & 0.994  & 0.0004  & $^{+0.00038}_{-0.00029}$ \\
\end{tabular}
\caption{Summary of the results of the measurement of the \pt\ distribution
of the \zb\ boson. The range of \pt\ corresponds to the intervals used for
binning the data. The product of the acceptance and efficiency is given as
$\epsilon\cdot a$, $b$ is the estimated number of background events, the
measured differential cross section is ${d\sigma^\prime/dp_T}$, the
correction for resolution-smearing is specified by $\alpha(p_T)$, and the 
corrected differential cross section is specified by ${d\sigma/dp_T}$. The 
uncertainty on the differential cross section includes both systematic and
statistical uncertatinties, but does not include overall normalization
uncertainty due to the luminosity of $\pm 4.4$\%.}
\label{tab:prd_results}
\end{center}
\end{table}

\twocolumn

\onecolumn

\twocolumn

\begin{figure}[htpb!]
  \centerline{\psfig{figure=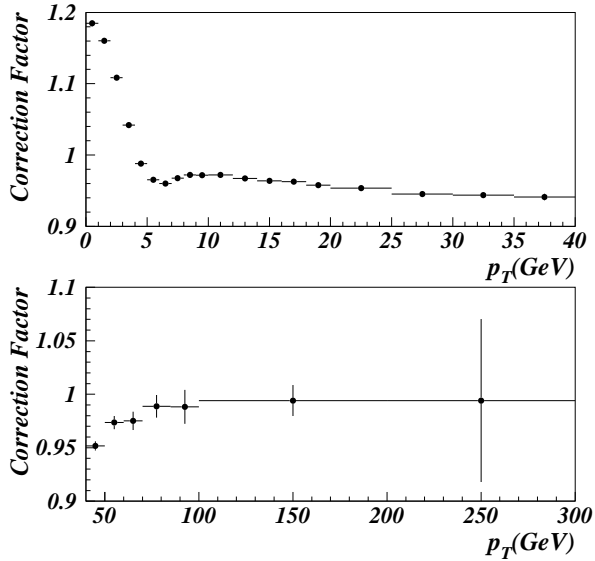,width=3.2in}}
  \caption{Smearing correction factor $\alpha(p_T)$ as
  a function of \pt.}
  \label{fig:smear_corr}
\end{figure}

\section{Smearing Corrections}
\label{sec-smearcorr}
The results shown in Fig.~\ref{fig:meas_results} still contain the
residual effects of detector smearing. We correct the 
measured
cross section for the effects of detector smearing using the ratio
of generated to resolution-smeared ansatz \pt\ distributions:
\begin{equation}\label{eq:alphasmear}
  \alpha(p_T) =
  {F(p_T;g_2)\over \int dp_T\ R(p_T,p^\prime_T)F(p_T;g_2)}
\end{equation}
where $p^\prime_T$ is the smeared value of \pt, 
$\alpha(p_T)$ is the correction factor, $F(p_T;g_2)$
is the ansatz function with parameter $g_2$ and
$R(p_T,p^\prime_T)$ is the resolution function.

As the ansatz function, we use the calculation from
\LEGA\ fixing $g_1=0.11$ \gevsq\ and $g_3=-1.5$ \gev$^{-1}$. 
We use $g_2=0.59$ \gevsq\ for our central value.

Figure~\ref{fig:smear_corr} shows the smearing correction as a
function of \pt. The largest effect occurs at low-\pt\ where the
smearing causes the largest fractional change in \pt\ and where 
the kinematic boundary at \pt$=0$ results in non-Gaussian 
smearing---the \pt\ is preferentially increased by the smearing rather than 
being a symmetric effect. Table~\ref{tab:prd_results} includes the 
value of the smearing correction for each bin of \pt.

\begin{figure}[htpb!]
  \centerline{\psfig{figure=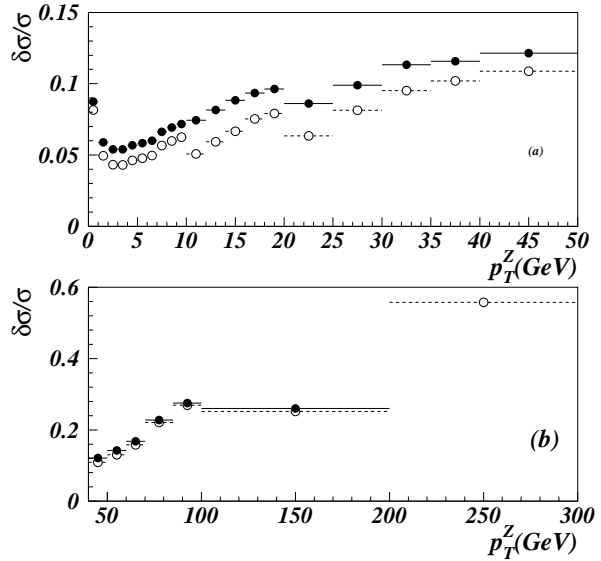,width=3.2in}}
  \caption{Fractional uncertainty in the cross section as a function
  of \pt\ including the statistical and systematic uncertainties (closed
  circles) and including only statistical uncertainties
  (open circles) for (a) $p_T<50$ \gevc\ and (b) $p_T>50$ \gevc.}
  \label{fig:fracerr}
\end{figure}

It is important that the smearing correction be insensitive to
significant variations in the ansatz function used to generate the
correction. We examine this issue by varying the parameter
$g_2=0.59\pm 0.06$\gevsq\ in the ansatz function by $\pm
1\sigma$, obtaining a variation of $<$1\% for all values of
\pt. For this variation in the parameter, the ansatz function
varies by $\approx$10\%. It is useful to compare
 the level of
uncertainty in the smearing correction to other components of
uncertainty in the measurement. Figure~\ref{fig:fracerr} shows the
fractional uncertainty on the differential cross section as a function
of \pt. Both the total uncertainties (in which systematic
uncertainties on the background, efficiency, and acceptance are
included with the statistical uncertainty), and the statistical
uncertainties alone are shown. The variations in the smearing
correction are at least a factor of five smaller than the other
uncertainties and therefore can be ignored.

The uncertainty in the smearing correction is also affected by the
uncertainty on the values of the resolutions used to generate the
smearing. We examine this uncertainty by varying the detector
resolutions by $\pm 1$ standard deviation from the nominal values.
Again, the effect on the smearing correction is negligible relative
to the other uncertainties in the measurement and this source of
uncertainty has been ignored.

\begin{figure}[htpb!]
  \psfig{figure=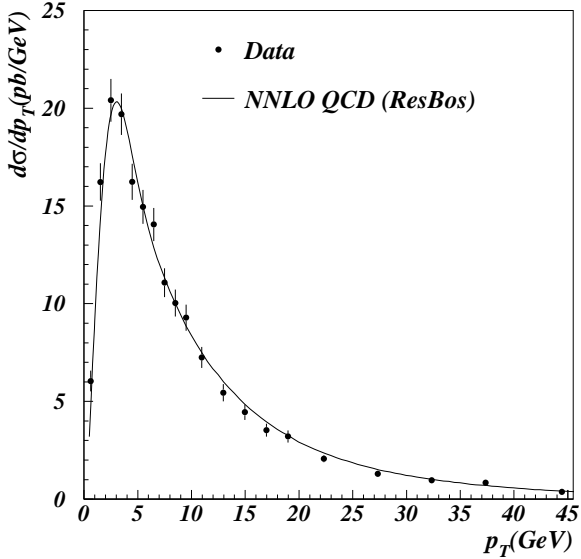,width=3.2in}
  \caption{Plot of the differential cross section (circles) as a function of
  \pt. The line is the result of the NNLO calculation from \RESBO\
  using the CTEQ4M pdf. Theory has been normalized to the
  data.}
  \label{fig:dsdpt_resum_lin}
\end{figure}

\section{Results}
\label{sec-results}

Table \ref{tab:prd_results} shows the final numerical results for
the measurement of \dsdpt\ using a total of 6407
events. The uncertainties on the data points include statistical
and systematic contributions. There is an additional normalization
uncertainty of $\pm$4.4\%\ from the uncertainty on the integrated
luminosity~\cite{wzcross} that
is included in neither the plots nor the table, but must be taken
into account in any fits requiring an absolute cross section.

\begin{figure}[htpb!]
  \centerline{\psfig{figure=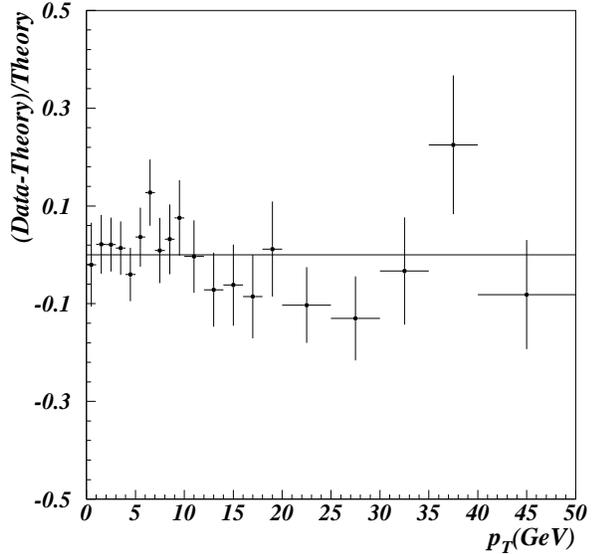,width=3.2in}}
  \caption{Fractional difference between the data and the NNLO
  calculation from \RESBO\ for the differential cross section as
  a function of \pt.}
  \label{fig:dtt_lowpt_resum}
\end{figure}

\begin{figure}[htpb!]
  \centerline{\psfig{figure=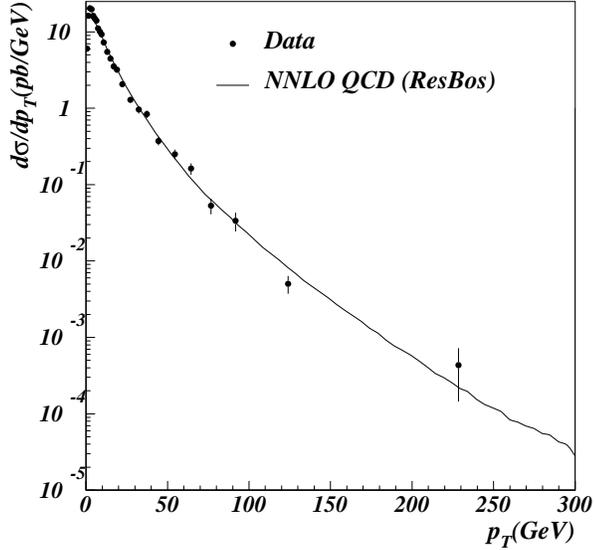,width=3.2in}}
  \caption{Plot of the differential cross section (circles) as a function of
  \pt. The line is from the NNLO calculation from \RESBO\ using CTEQ4M as
  the pdf. Theory has been normalized to the data.}
  \label{fig:dsdpt_resum_log}
\end{figure}

\begin{figure}[t]
  \centerline{\psfig{figure=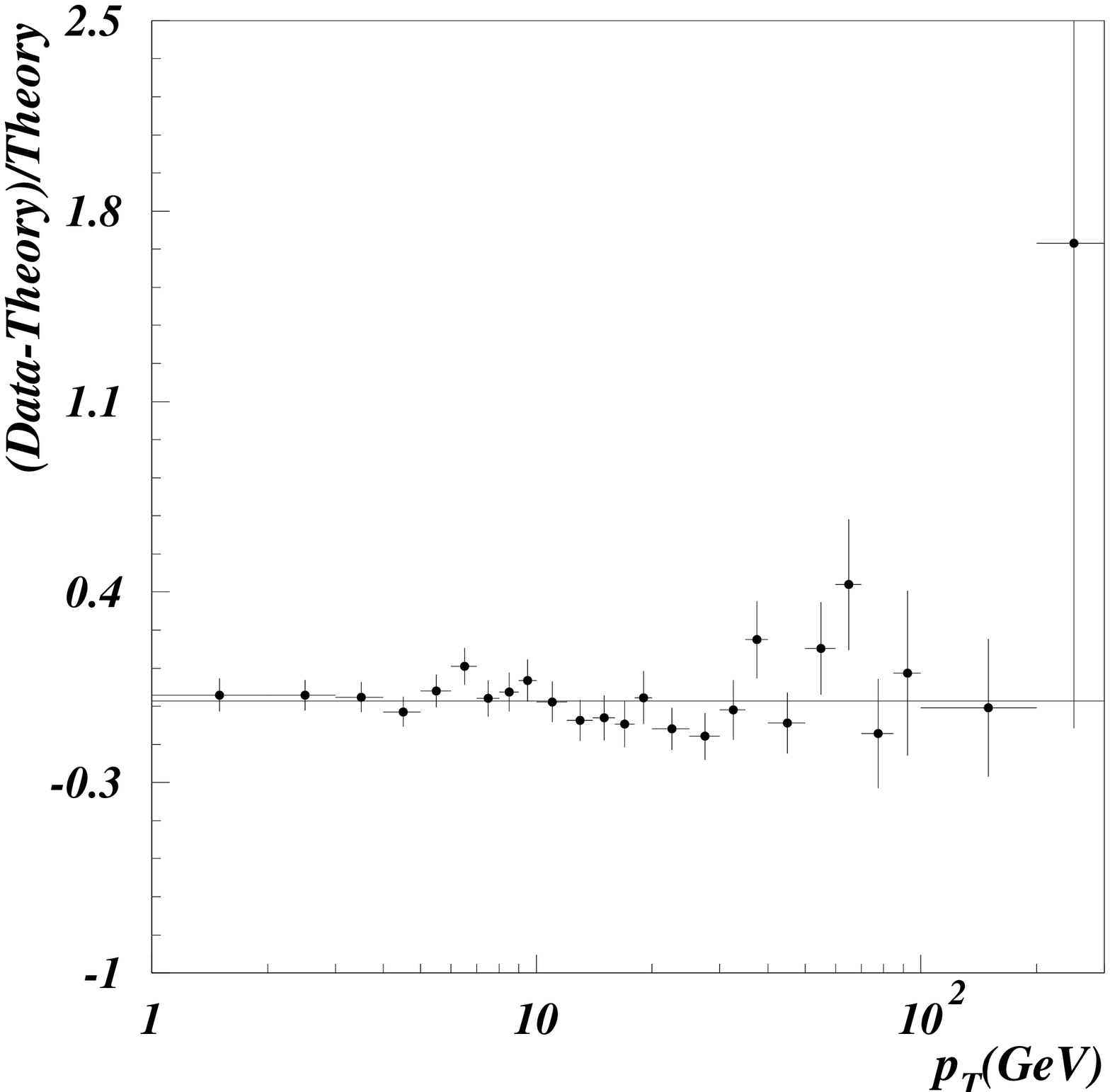,width=3.2in}}
  \caption{Fractional difference between the data and the NNLO
  calculation from \RESBO\ for the differential cross section as
  a function of \pt.}
  \label{fig:dtt_allpt_resum}
\end{figure}

Figures~\ref{fig:dsdpt_resum_lin}--\ref{fig:dsdpt_resum_log} and 
\ref{fig:dtt_allpt_resum} show the
final, smearing-corrected \pt\ distribution compared to the fully
resummed calculation as calculated by \RESBO. The calculation uses 
the CTEQ4M pdf's and the Ladinsky-Yuan parameterization for the 
non-perturbative function with the published values for the
$g_i$ parameters: $g_1=0.11$ \gevsq, $g_2=0.58$ \gevsq, 
and $g_3=-1.5$ \gev$^{-1}$. The data points in Figs.~\ref{fig:dsdpt_resum_lin}
and \ref{fig:dsdpt_resum_log} are placed at the average
\pt\ of the bin, as given by theory, rather than
at the center of the bin. (Only the first and twenty-fifth bins
change appreciably, $-24$\% and $-17$\% respectively, relative to the
bin center.) For the resummed calculation, the total
cross section predicted by the theory (220 pb) has been
normalized to the data. A $\chi^2$ comparison of the prediction
to the data yields a $\chi^2/$d.o.f.$=26.7/25$. We observe good
agreement with the fully resummed calculation for all values of \pt.

\begin{figure}[b]
  \centerline{\psfig{figure=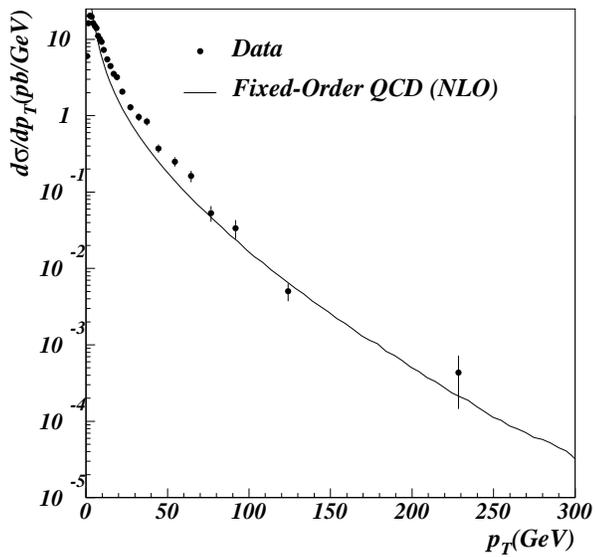,width=3.2in}}
  \caption{Plot of the differential cross section (circles) as a function of
  \pt. The line is the result of the fixed-order (NLO) calculation of
  the differential cross section.}
  \label{fig:dsdpt_nlo_log}
\end{figure}

\begin{figure}[b]
  \centerline{\psfig{figure=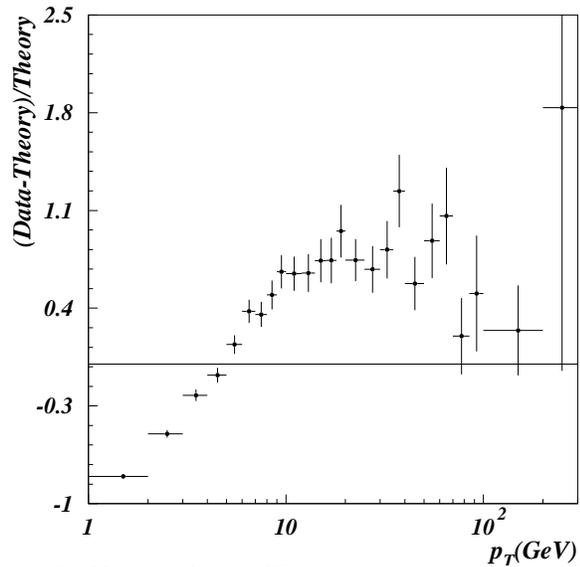,width=3.2in}}
  \caption{Fractional difference between data and the fixed-order
  calculation from \RESBO\ for the differential cross section as
  a function of \pt.}
  \label{fig:dtt_hipt_nlo}
\end{figure}

Figures~\ref{fig:dsdpt_nlo_log} and \ref{fig:dtt_hipt_nlo} compare the
data to the fixed-order (NLO) perturbative calculation as calculated by
\RESBO\ and using the CTEQ4M pdf's. We observe strong disagreement at 
low-\pt, as expected due to the divergence of the NLO calculation at 
\pt$=0$, and a
significant enhancement of the cross section relative to the
prediction at moderate values of \pt, confirming the increase in
the cross section from soft gluon emission.

\section{Conclusions}
\label{sec-conclusions}

We have measured the differential cross section as a function of the
transverse momentum of the \zb\ boson. Fitting for the
value of the non-perturbative parameter $g_2$, we obtain
$g_2=0.59\pm0.06$ \gevsq, which is significantly
more precise than previous determinations. We observe good agreement 
between the measurement and the resummation calculations for all values of \pt.

\section*{ Acknowledgements }
%
We thank the Fermilab and collaborating institution staffs for
contributions to this work and acknowledge support from the 
Department of Energy and National Science Foundation (USA),  
Commissariat  \` a L'Energie Atomique (France), 
Ministry for Science and Technology and Ministry for Atomic 
   Energy (Russia),
CAPES and CNPq (Brazil),
Departments of Atomic Energy and Science and Education (India),
Colciencias (Colombia),
CONACyT (Mexico),
Ministry of Education and KOSEF (Korea),
and CONICET and UBACyT (Argentina).

\end{document}